\documentclass[]{aa}  

\usepackage{graphicx}
\usepackage{txfonts}
\usepackage[breaklinks=true]{hyperref}
\hypersetup{
    colorlinks=true,
    citecolor=blue,
    linkcolor=blue,
    filecolor=magenta,      
    urlcolor=cyan,
    }

\usepackage{natbib,twoopt}
\usepackage[breaklinks=true]{hyperref} 
\bibpunct{(}{)}{;}{a}{}{,} 
\makeatletter
\newcommandtwoopt{\citeads}[3][][]{\href{http://adsabs.harvard.edu/abs/#3}%
{\def\hyper@linkstart##1##2{}%
\let\hyper@linkend\@empty\citealp[#1][#2]{#3}}}
\newcommandtwoopt{\citepads}[3][][]{\href{http://adsabs.harvard.edu/abs/#3}%
{\def\hyper@linkstart##1##2{}%
\let\hyper@linkend\@empty\citep[#1][#2]{#3}}}
\newcommandtwoopt{\citetads}[3][][]{\href{http://adsabs.harvard.edu/abs/#3}%
{\def\hyper@linkstart##1##2{}%
\let\hyper@linkend\@empty\citet[#1][#2]{#3}}}
\newcommandtwoopt{\citeyearads}[3][][]%
{\href{http://adsabs.harvard.edu/abs/#3}
{\def\hyper@linkstart##1##2{}%
\let\hyper@linkend\@empty\citeyear[#1][#2]{#3}}}
\makeatother
\usepackage{placeins}
\usepackage{acronym}

\acrodef{whim}[WHIM]{warm-hot intergalactic medium}
\acrodef{cgm}[CGM]{circumgalactic medium}
\acrodef{srg}[\emph{SRG}]{\emph{Spektrum Roentgen
Gamma}}
\acrodef{erosita}[eROSITA]{extended ROentgen Survey with an Imaging Telescope Array}
\acrodef{erass}[eRASS]{\ac{erosita} All-Sky Survey}
\acrodef{ls}[LS]{DESI Legacy Surveys}
\acrodef{rass}[RASS]{ROSAT All-Sky Survey}
\acrodef{efeds}[eFEDS]{\ac{erosita} Final Equatorial-Depth Survey}
\acrodef{sdss}[SDSS]{Sloan Digital Sky Survey}
\acrodef{srg}[SRG]{Spectrum-Roentgen-Gamma}
\acrodef{agn}[AGN]{active galactic nucleus}
\acrodef{xrb}[XRB]{X-ray binary}
\acrodefplural{xrb}[XRBs]{X-ray binaries}
\acrodefplural{agn}[AGNs]{active galactic nuclei}
\acrodef{arf}[ARF]{ancillary response file}
\acrodef{rmf}[RMF]{response matrix file}
\acrodef{psf}[PSF]{point spread function}
\acrodef{cie}[CIE]{collisional ionization equilibrium}
\acrodef{dem}[DEM]{differential emission measure}
\acrodef{gdem}[GDEM]{Gaussian differential emission measure}
\acrodef{tm}[TM]{telescope module}
\acrodef{mcmc}[MCMC]{Markov chain Monte Carlo}
\acrodef{apec}[APEC]{Astrophysical Plasma Emission Code}
\acrodef{tbabs}[TBABS]{Tuebingen-Boulder ISM absorption model}
\acrodef{fwhm}[FWHM]{full width at half maximum}
\acrodef{sz}[SZ]{Sunyaev–Zeldovich}
\acrodef{lrg}[LRG]{luminous red galaxy}
\acrodef{ccd}[CCD]{charge-coupled device}
\acrodef{smf}[SMF]{stellar mass function}
\acrodef{sfr}[SFR]{star forming rate}
\acrodef{mcmc}[MCMC]{Markov chain Monte Carlo}
\acrodef{cmb}[CMB]{cosmic microwave background}
\acrodefplural{cmb}[CMB \citealt{Planck15, Planck2020}]{cosmic microwave background}
\acrodef{bbn}[BBN]{big bang nucleosynthesis}
\acrodef{cxb}[CXB]{cosmic X-ray background}
\acrodef{frb}[FRB]{fast radio burst}
\acrodef{1t}[1-T]{single temperature}
\acrodef{pi}[PI]{pulse invariant}
\acrodef{nyu-vagc}[NYU-VAGC]{New York University Value Added Galaxy Catalogue}
\acrodef{uvb}[UVB]{cosmic ultraviolet background}

\begin{document}

   \title{The SRG/eROSITA all-sky survey}

   \subtitle{X-ray emission from the warm-hot phase gas in long cosmic filaments}

   \author{X.~Zhang\inst{\ref*{inst:mpe}}\and
          E.~Bulbul\inst{\ref*{inst:mpe}}\and
          N.~Malavasi\inst{\ref*{inst:mpe}}\and
          V.~Ghirardini\inst{\ref*{inst:mpe},\ref*{inst:inaf-oas}}\and
          J.~Comparat\inst{\ref*{inst:mpe}}\and
          M.~Kluge\inst{\ref*{inst:mpe}}\and
          A.~Liu\inst{\ref*{inst:mpe}}\and
          A.~Merloni\inst{\ref*{inst:mpe}}\and
          Y.~Zhang\inst{\ref*{inst:mpe}}\and
          Y.~E.~Bahar\inst{\ref*{inst:mpe}}\and
          E.~Artis\inst{\ref*{inst:mpe}}\and
          J.~S.~Sanders\inst{\ref*{inst:mpe}}\and
          C.~Garrel\inst{\ref*{inst:mpe}}\and
          F.~Balzer\inst{\ref*{inst:mpe}}\and
          M.~Brüggen\inst{\ref*{inst:hamburg}}\and
          M.~Freyberg\inst{\ref*{inst:mpe}}\and
          E.~Gatuzz\inst{\ref*{inst:mpe}}\and
          S.~Grandis\inst{\ref*{inst:innsbruck}}\and
          S.~Krippendorf\inst{\ref*{inst:lmu},\ref*{inst:asc}}\and
          K.~Nandra\inst{\ref*{inst:mpe}}\and
          G.~Ponti\inst{\ref*{inst:inaf-oab},\ref*{inst:mpe}}\and
          M.~Ramos-Ceja\inst{\ref*{inst:mpe}}\and
          P.~Predehl\inst{\ref*{inst:mpe}}\and
          T.~H.~Reiprich\inst{\ref*{inst:aifa}}\and
          A.~Veronica\inst{\ref*{inst:aifa}}\and
          M.~C.~H.~Yeung\inst{\ref*{inst:mpe}}\and
          S.~Zelmer\inst{\ref*{inst:mpe}}
          }

   \institute{Max-Planck-Institut fur Extraterrestrische Physik, Giessenbachstrasse, 85748, Garching, Germany \label{inst:mpe}\\
              \email{xzhang@mpe.mpg.de}
         \and
            INAF, Osservatorio di Astrofisica e Scienza dello Spazio, via Piero Gobetti 93/3, 40129 Bologna, Italy \label{inst:inaf-oas}
         \and
            Hamburger Sternwarte, Gojenbergsweg 112, 21029, Hamburg, Germany \label{inst:hamburg}
         \and
            University Observatory Munich, Faculty of Physics, Ludwig-Maximilians-Universität, Scheinerstr. 1, 81679, Munich, Germany \label{inst:lmu}
        \and
            Universität Innsbruck, Institut für Astro- und Teilchenphysik, Technikerstrasse 25/8, 6020, Innsbruck, Austria \label{inst:innsbruck}
        \and
            Arnold Sommerfeld Center for Theoretical Physics, LMU Munich, Theresienstr. 37, 80333 Munich, Germany \label{inst:asc}
        \and
            INAF, Osservatorio Astronomico di Brera, Via E. Bianchi 46, 23807, Merate, (LC), Italy \label{inst:inaf-oab}
        \and
            Argelander Institute for Astronomy, University of Bonn, Auf dem Hügel 71, 53121, Bonn, Germany \label{inst:aifa}
             }

   \date{}


  \abstract{
    The properties of the warm-hot intergalactic medium (WHIM) in cosmic filaments are among the least quantified units in modern astrophysics. The Spectrum Roentgen Gamma/eROSITA All Sky Survey ((SRG/eRASS) provides a unique opportunity to study the X-ray emission of the WHIM. 
    We applied both imaging and spectroscopic stacking techniques to the data of the first four eRASS scans to inspect the X-ray emissions from 7817 cosmic filaments identified from Sloan Digital Sky Survey (SDSS) optical galaxy samples. We obtained a $9\sigma$ significant detection of the total X-ray signal from filaments in the 0.3--1.2~keV band. Here, we introduce a novel method to 
    estimate the contamination fraction from unmasked X-ray halos, active galactic nuclei, and X-ray binaries associated with filament galaxies. We found an approximately 40\% contamination fraction for these unmasked sources, suggesting that the remaining 60\% of the signal could be coming from the WHIM and a $5.4\sigma$ detection significance of the WHIM. Moreover, we modeled the temperature and baryon density contrast of the detected WHIM by fitting the stacked spectrum and surface brightness profile. The best-fit temperature $\log(T/\mathrm{K})=6.84\pm0.07$, obtained by using a single temperature model, is marginally higher than in the simulation results. This could be due to the fitting of a single temperature model on a multi-temperature spectrum. Assuming a 0.2 solar abundance, the best-fit baryon density contrast $\log\Delta_\mathrm{b}=1.88\pm0.18$ is in general agreement with the X-ray emitting phases in the IllustrisTNG simulation. This result suggests that the broadband X-ray emission traces the high end of the temperature and density values that characterize the entire WHIM population.
    
  }

\titlerunning{X-ray emission from cosmic filaments and WHIM}
\maketitle

\section{Introduction}

The \ac{whim} permeating cosmic filaments, with a temperature of $10^{5-7}$~K and density of $n_\mathrm{H}<10^{-4}$~cm$^{-3}$, makes up the majority of the ``missing baryon'' in the low-redshift Universe predicted by numerical simulations \citep[e.g.,][]{Fukugita1998,Cen1999,Shull2012,Tuominen2021}. Recent measurements on the dispersions of localized \acp{frb} that trace the total line-of-sight free electron column density quantify the cosmic baryon density in the low-redshift Universe in a new way \citep{Macquart2020, Yang2022, Wang2023}. These \ac{frb} cosmic baryon density measurements are consistent with the results from other probes, such as \acp{cmb} and big bang nucleosynthesis \citep{Pitrou2018}, which conclude that baryons are not missing in the low-redshift Universe. However, the \ac{frb} dispersion measure technique cannot provide additional detailed information on the baryons, such as density and temperature. Exploring these properties using other observation techniques, such as X-ray observations and \ac{sz} effects, remains one of the primary investigation goals in modern astrophysics \citep{Driver2021}.

Currently, absorption studies by using OVI lines in the far ultraviolet bands can successfully trace the $10^{5.5}$~K low-temperature phases \ac{whim} \citep{Danforth2016}. For X-ray absorption lines that trace $\sim10^6$~K phases \ac{whim}, while there are many cases of plausible detections \citep{Nicastro2005,Fang2010,Nicastro2018}, these constraints suffer from potential contamination from other absorption sources \citep[e.g.][]{Johnson2019,DorigoJones2022,Gatuzz2023}. This method is also currently restricted to a handful of bright background blazars (see \citealt{Fang2022} for a review). Meanwhile, X-ray emission studies have been mostly focused on the brightest part of the \ac{whim} with a relatively  high density contrast, which resides in bridges connecting galaxy cluster pairs or along the filaments \citep{Werner2008, Bulbul2016, Akamatsu2017,Reiprich2021} or circumcluster space of massive galaxy clusters \citep{Eckert2015,Bulbul2016,Reiprich2021}. Similarly, the number of these systems is limited, and there is less information about the properties of the \ac{whim} in the environments of long cosmic filaments. 

Wide-area optical spectroscopic surveys have mapped the large-scale galaxy distributions up to redshifts of 0.7 \citep{Reid2016}, thus offering opportunities for a complete examination of the \ac{whim} emission in the vast space of the cosmic filaments. Two pilot studies have shown mild positive detections ($4.2\sigma$ and $3.8\sigma$) of stacked X-ray surface brightness emission from cosmic filaments using \ac{rass} and \ac{efeds} data \citep{Tanimura2020-xray,Tanimura2022}, demonstrating the potential of X-ray stacking analysis to trace the emission in cosmic filaments. Similarly, mild positive detections have also been reported by stacking the \ac{sz} effect signal that traces the pressure of the intergalactic medium \citep{deGraaff2019,Tanimura2019,Tanimura2020-sz}. However, in these previous works, detections in the very soft X-ray band $<0.6$~keV at the observed frame, needed to constrain the temperature of the redshifted $T<10^7$~K gas, are still lacking. Meanwhile, the energy resolution was limited in the observer frame due to the broad redshift distribution of the filament sample, limiting the ability to measure the plasma temperature in the soft band. Moreover, the pilot X-ray image stacking studies did not account for the signal from unmasked X-ray sources, which could contribute a significant fraction of the total emission. In this work, we improve upon the previous stacking analyses by extending the bandpass for spectral analysis, increasing the energy resolution by using a rest-frame spectral stacking scheme, accounting for the emission from unmasked low-mass halos and point sources and utilizing the new \ac{erass} data. This gives us access to an extensive area with an unprecedented depths, with excellent soft band sensitivity that is essential for detecting the faint X-ray emission from WHIM gas. 

In this work, we assume a fiducial Lambda cold dark matter cosmology with parameters: $H_0=70$ km s$^{-1}$ Mpc$^{-1}$, $\Omega_\mathrm{m}=0.3$, $\Omega_\Lambda=0.7$. We adopt the ratio between baryon density parameter to matter density parameter from the \emph{Planck}18 cosmology results, namely: $\Omega_\mathrm{b}/\Omega_\mathrm{m}=0.158$ \citep{Planck2020}. At $z=0$, the mean baryon density is $\overline{\rho}_\mathrm{b}(0)=4.51\times10^{-31}$ g cm$^{-3}$. The baryon density contrast is defined as $\Delta_\mathrm{b}\equiv\rho_\mathrm{b}/\overline{\rho}_\mathrm{b}$. Throughout this article, $\log$ refers to the logarithm to the base 10, and $\ln$ refers to the natural logarithm. Unless otherwise indicated, all sizes, scales, and lengths are given in physical (proper) scale. 
This article is organized as follows. In Sect.~\ref{sect:obs}, we describe the X-ray data products and the optical filament catalog we used. In Sect.~\ref{sect:profile_stacking}, we present the surface brightness profile stacking method and results. In Sect.~\ref{sect:contam}, we estimate the fraction of detected signal due to unmasked galactic sources. Section~\ref{sect:spec} presents the spectral stacking method and temperature measurements. The estimation of the detected X-ray emitting phases and the comparison with numerical simulations are presented in Sect. \ref{sect:phase}. We conclude our work in Sect. \ref{sect:summary}.

\section{Observations, optical filament catalog, and data preparation}\label{sect:obs}

The stacked first four scans of the eROSITA All-Sky Survey (hereafter eRASS:4) data were collected from  December 12, 2019 to December 19, 2021. The data were processed with eSASS \citep{Brunner2022} pipeline version 020, which is the same as the version 010 used for eRASS1 data release with improvements on boresight correction, detector noise suppression, and subpixel position computation \citep{Merloni2024}. We only selected events from \acp{tm} 1, 2, 3, 4, and 6 (hereafter, TM8) to avoid the systematic uncertainties caused by the optical light leak in \acp{tm} 5 and 7 \citep{Predehl2021}. We used the tools in the software package eSASS version \texttt{eSASSusers\_211214\_0\_4} to generate \ac{erass} data products.

\begin{figure}
    \centering
    \includegraphics[width=0.5\textwidth]{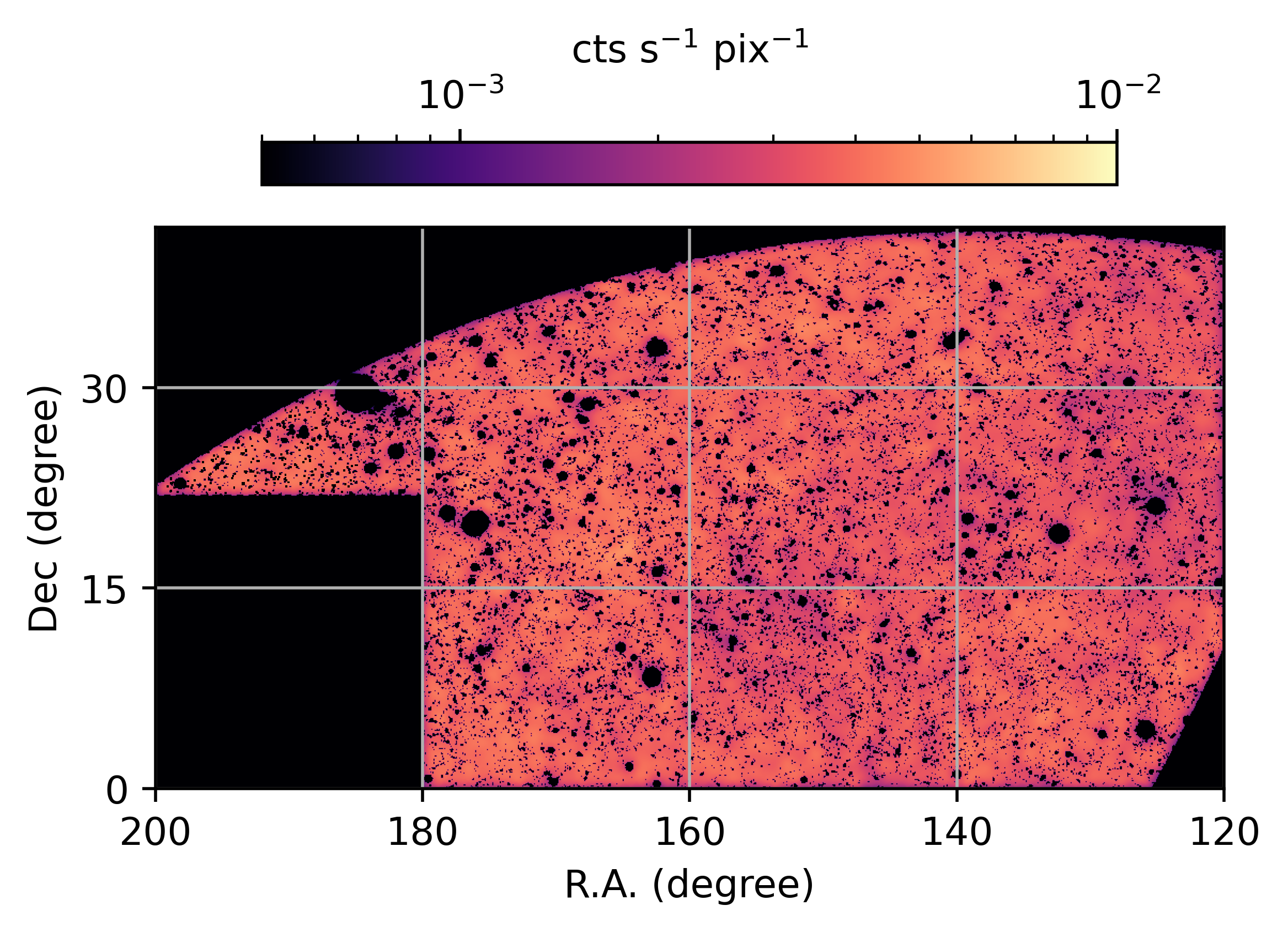}
    \caption{Smoothed 0.3--1.2~keV eROSITA count rate map of the analysis footprint. The missing pixels show source-masked regions.}
    \label{fig:footprint}
\end{figure}

\begin{figure}
    \centering
    \includegraphics[width=0.45\textwidth]{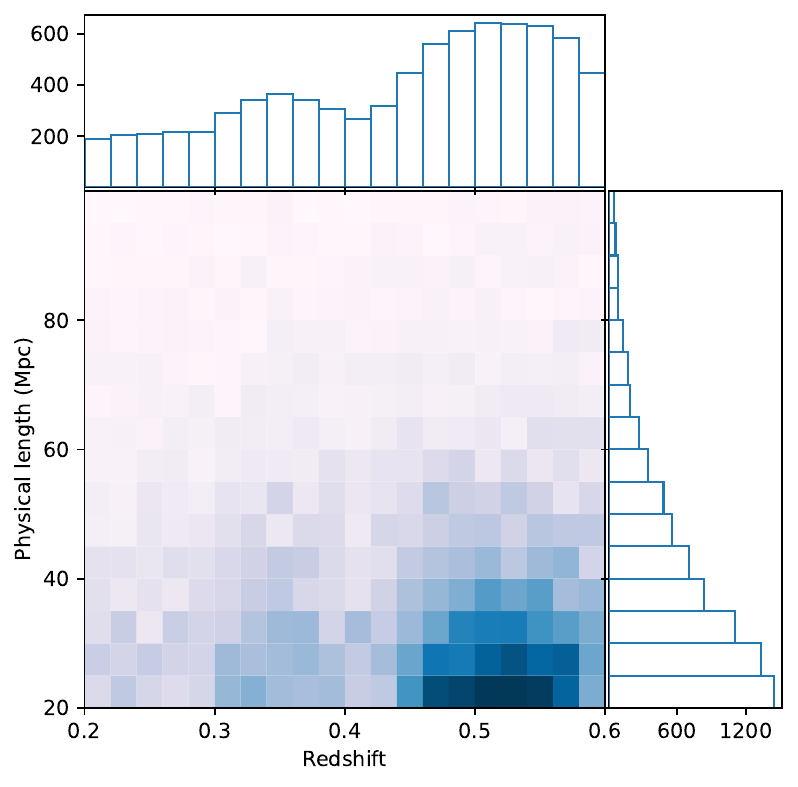}
    \caption{Distribution of the selected filaments in the redshift and physical length space.}
    \label{fig:l-z_hist}
\end{figure}

We adopted an optical filament catalog \citep{Malavasi2020}, compiled using the \texttt{DisPerSE} software \citep{Sousbie2011a,Sousbie2011b} and based on the \ac{sdss} LOWZ+CMASS galaxy catalogs \citep{Reid2016}. The filament catalog contains 63391 filaments after a $3\sigma$ persistence cut without any smoothing applied on the galaxy density field. The original filament catalog is stored in the comoving frame and Cartesian coordinates. Therefore, we converted the coordinates back to (right ascension, declination, redshift) using the conversion functions and the cosmology used in \citet{Malavasi2020}. The physical length of the filaments was calculated using the assumed cosmological parameters above. The cross-section of the SDSS LOWZ+CMASS footprint \citep{Reid2016} and the western Galactic hemisphere of the eROSITA All-Sky Survey determines a $\sim3200\deg^2$ common analysis footprint. In addition, the bright foreground regions of the eROSITA Bubble \citep{Predehl2020}, the Virgo Cluster \citep{Boehringer1994,McCall2024}, and the Galactic plane $|b|<20^\circ$ were excluded from the analysis for simplicity. The total analysis footprint before source masking has a solid angle is shown in Fig.~\ref{fig:footprint},  amounting to 2275~deg$^2$. In this footprint, we select 7817 filaments in a redshift range $0.2<z<0.6$ and a physical length range 20~Mpc~$<l_\mathrm{fil}<$~100~Mpc for the stacking analysis (see Fig.~\ref{fig:l-z_hist}). 

\subsection{Source masking}\label{sect:mask}

We masked out all eRASS:4 X-ray sources in the 0.2--2.3~keV eSASS detection catalog\footnote{Proprietary eROSITA data. Version \texttt{all\_s4\_SourceCat1B\_221031\_poscorr\_mpe\_photom}}. The detection pipeline used in this work is the same as the version used for the first release of the All-Sky Survey data \citep{Merloni2024}. The catalog contains point-like and extended sources, such as supernova remnants, galaxies, galaxy groups, and clusters. To remain conservative, we did not apply selection cuts to the contaminating X-ray emitting sources.

We used a background level-dependent radius to exclude detected sources as follows. 
For each source, the surface brightness profile can be approximately described as 

\begin{equation}
    S(r) = S_0\left(1+\frac{r^2}{r_\mathrm{PSF}^2 + r_\mathrm{EXT}^2}\right)^{-1.5},
\end{equation}
\noindent where $r_\mathrm{PSF}=9''$ is the survey averaged \ac{psf} core radius parameter obtained from point source stacking and $r_\mathrm{EXT}$ is the parameter \texttt{EXT} given by the catalog if the source is extended. We adopted the radius at which the source profile reaches 0.1 (10\%) of the background level as the masking radius, namely, $S(r_\mathrm{cut}) = 0.1 \times S_\mathrm{bkg}$, where $S_\mathrm{bkg}$ is obtained from \texttt{ML\_BKG\_1} from the detection pipeline.  

For extended sources with large angular sizes, the eSASS detection pipeline usually returns several detections known as split sources \citep{Liu2022,Liu2024}. These large angular size sources are already included in the eRASS1 galaxy cluster and group catalog \citep{Bulbul2024}. Therefore, we masked out all sources in the eRASS1 galaxy cluster and group catalog with radii of $1.5\times r_{500}$. Additionally, we adopted an optical cluster catalog compiled by the redMaPPer algorithm \citep{Rykoff2014} on the DESI Legacy Survey \citep{Dey2019} DR9 + DR10 \textit{grz} images in a blind mode \citep{Kluge2024}. We excluded sources with a richness of $\lambda>20$ and redshift  of $0.2<z<0.8$ at radii of $1.5\times r_\lambda$, where $r_\lambda = (\lambda/100)^{1/5}$ $h^{-1}$~Mpc (this is the richness scaled radius from the redMaPPer algorithm). This cut safely removes clusters of galaxies with high completeness and high purity, especially for high-redshift systems whose X-ray luminosity is below the eRASS1 detection limit.

To conservatively remove the emission from low-richness galaxy groups in the local Universe, we also adopted a galaxy group catalog \citep{Tinker2021} based on the SDSS Main Galaxy Sample \citep{Strauss2002}. We exclude sources with $n_\mathrm{sat}>1$ in radii of $1.5\times r_{500}$. If a source displays $1.5\times r_{500}<400$~kpc, we used a fixed value of 400~kpc instead. In addition to those cluster and group catalogs, we masked out critical points identified by DisPerSE of type maximum and bifurcation in the filament catalog in a redshift range of $0.2<z<0.6$ with fixed radii of 2~Mpc.  This range is what we consider  as the nodes of the large-scale structure \citep[see][for details]{Malavasi2020}. Details on all masked sources are listed in table \ref{tab:source_mask}.

We note that we did not mask optically selected clusters with $\lambda<20$, which is the richness range where the identified clusters or groups are usually seen to suffer from significantly high contamination due to projection effects \citep{Costanzi2019,Grandis2021}. Masking these sources out will significantly reduce the useful solid angle in the analysis footprint. Given that the $\lambda<20$ halos and the unmasked sources below the X-ray detection limit also contribute to the stacked signal in this work (due to their associations with the large-scale structure of the Universe), we estimated the fraction of these unmasked X-ray sources and model it out from our final analysis when determining the significance of the emission from the \ac{whim} (see Sect.~\ref{sect:contam}).

\begin{table*}[]
\caption{Summary of the masked sources.}\label{tab:source_mask}
    \centering
    \begin{tabular}{cccc}
        \hline\hline
        Catalog & Type & Selection & Masking radius   \\
        \hline
        eRASS:4 0.2--2.3~keV detection & X-ray & all & $r|_{S=0.1S_\mathrm{bkg}}$\\
        eRASS1 cluster \& group & X-ray & all & $1.5\times r_\mathrm{500c}$ \\
        redMaPPer on LS DR9 \& DR10 & Optical & $\lambda>20$, $0.2<z<0.8$ & $1.5\times r_\lambda$ \\
        SDSS MGS group & Optical & $n_\mathrm{sat}>1$& $\max\{400\mathrm{kpc}, 1.5\times r_\mathrm{500c}\}$  \\
        Critical points in filament catalog & DisPerSE & maxima + bifurcations & 2 Mpc\\
        \hline
    \end{tabular}
\end{table*}
\subsection{Healpix map creation}\label{sect:healpix}
We adopted the 0.3--1.2~keV band in the imaging analysis. This band is optimized for maximizing the \ac{whim} emission from our selected filament sample with $z\sim0.47$ (redshifted), assuming a $kT<1$~keV gas temperature.
We split the analysis footprint into 27 tiles, each size of 10~deg~$\times$~10~deg. For each tile, we merged the event list using \texttt{evtool} and obtain flare-free time interval using \texttt{flaregti} with parameters \texttt{pimin=5000}, \texttt{threshold=1.1}, and \texttt{source\_size=150}. We used \texttt{evtool} and criteria 
\texttt{PATTERN=15}, \texttt{FLAG=0xE000F000}, and \texttt{GTI=FLAREGTI} to select events in the flare-free time interval and generate count images with a pixel size of $16''$. We used \texttt{expmap} to create a corresponding vignetting corrected exposure map for each count image. After masking the sources, we binned the tile image pixels to NSIDE=1024 HEALPix\footnote{\url{http://healpix.sourceforge.net}} \citep{Gorski2005,Zonca2019} map pixels. The NSIDE=1024 HEALPix has a $3.4'$ pixel size corresponding to 1.4 Mpc at $z=0.6$. 
For the count map, the count number of the $i$~th pixel is $C_{\mathrm{HP},i} = \sum_j C_{ij}$, where $C_{ij}$ is the count number of the $j$~th image pixel that is located in the $i$~th HEALPix pixel and is not masked. For the exposure map, the exposure time of the $i$~th pixel, $t_{\mathrm{HP},i} = (\sum_j t_{ij}\times \Omega_{ij})/{\Omega_{\mathrm{HP},i}}$,
where $\sum_j t_{ij}$ and $\Omega_{ij}$ are the exposure and solid angle of the $j$~th image pixel that is located in the $i$~th HEALPix pixel and is not masked, and $\Omega_{\mathrm{HP},i}$ is the solid angle of the $i$~th HEALPix pixel.

\section{Surface brightness stacking}\label{sect:profile_stacking}
We use $i$, $j$, and $k$ to denote the indices of pixel, filament, and radial transverse distance bin, respectively, and we have $n_\mathrm{HP}$ HEALPix pixels, $n_\mathrm{fil}$ filaments and $n_\mathrm{bin}$ distance bins. For each combination of the $i$~th pixel $\mathcal{P}_i$, the $j$~th filament $\mathcal{F}_j$ and the $k$~th distance bin $\mathcal{B}_k$, we assigned a Boolean coefficient $b_{ijk}$, whose value is 1 (or 0) if $\mathcal{P}_i$ is in (or not in) the $k$~th distance bin of $\mathcal{F}_j$. Therefore, for each filament, $\mathcal{F}_j$, the count number in the $k$th transverse distance bin is $C_{jk}=\sum_{i}^{n_\mathrm{HP}}{b_{ijk}\times C_{i}}$, and the exposure time in the $k$th transverse distance bin is $t_{jk}=\sum_{i}^{n_\mathrm{HP}}{b_{ijk}\times t_{i}}$. We used exposure-weighted averages to obtain the stacked surface brightness profile. The stacked profile, 

\begin{equation}\label{eq:individual_sb_prof-1}
    S_{\mathrm{X}}(r_k) = \frac{1}{\sum_{j}^{n_\mathrm{fil}}{t_{jk}}}\sum_{j}^{n_\mathrm{fil}}t_{jk}\times \frac{C_{jk}}{t_{jk}}=\frac{\sum_j^{n_\mathrm{fil}}C_{jk}}{\sum_j^{n_\mathrm{fil}}t_{jk}},
\end{equation}

\noindent is the quotient between the sum of the count profile and the sum of the exposure profile, where $r_k$ is the radial transverse distance of the $k$ th bin.  We used the averaged surface brightness in the outer part of the stacked profile as the local background, $S_\mathrm{X,bkg}$, and subtracted it from the entire profile to get a net surface brightness profile as

\begin{equation}\label{eq:individual_sb_prof-2}
S_\mathrm{X,net}(r_k)=S_{\mathrm{X}}(r_k)-S_\mathrm{X,bkg}.
\end{equation}

Due to the projection effects, the distance bins of different filaments at different redshifts could be simultaneously located at a single position. 
Therefore, the surface brightness in different bins is possibly correlated. We applied a bootstrapping method to construct resampled pixel sets to obtain the estimates of the mean and covariance of the stacked profile. We used $l$ to denote the index of the randomized pixel sample. The $l$th resample of the pixel set with the replacement is

\begin{equation}
    \boldsymbol{\mathcal{P}}_l^{'} = \left( \mathcal{P}_{l1}^{'}\quad \mathcal{P}_{l2}^{'}\quad \dots\quad \mathcal{P}_{ln_\mathrm{HP}}^{'} \right), \text{ with } \mathcal{P}_{li}^{'} \in \boldsymbol{\mathcal{P}}.
\end{equation}

Following Eqs. \ref{eq:individual_sb_prof-1} and \ref{eq:individual_sb_prof-2}, for each resampled pixel set $\boldsymbol{\mathcal{P}}_l^{'}$, we obtained a stacked surface brightness profile $\boldsymbol{S}_{\mathrm{X,net},l}^{'}(r_k)$.
After drawing $n_\mathrm{bs}$ random resamples, the bootstrapping estimate of the net surface brightness profile is 

\begin{equation}
    \hat{S}_\mathrm{X,net}(r_k)=\frac{1}{n_\mathrm{bs}-1}\sum_{l}^{n_\mathrm{bs}}S_{\mathrm{X,net},l}^{'}(r_k),
\end{equation}

\noindent and the bootstrapping estimate of the covariance between the $k_1$~th and $k_2$~th bin is 

\begin{align}\label{eq:cov_obs}
    \mathcal{C}_{k_1k_2}=&\frac{1}{n_\mathrm{bs}-1}\nonumber\\
    &\times\sum_{l}^{n_\mathrm{bs}}\left[S_{\mathrm{X,net},l}^{'}(k_1)-\hat{S}_\mathrm{X,net}(k_1)\right]\left[S_{\mathrm{X,net},l}^{'}(k_2)-\hat{S}_\mathrm{X,net}(k_2)\right].
\end{align}

\noindent The uncertainty of $\hat{S}_{\mathrm{X,net},k}$ can be taken from the diagonal element of the covariance matrix $\hat{\sigma}_{S_\mathrm{X},k}^2 = \mathcal{C}_{kk}$.

\subsection{Stacking the surface brightness profile}

\begin{figure}
    \centering
    \includegraphics[width=0.5\textwidth]{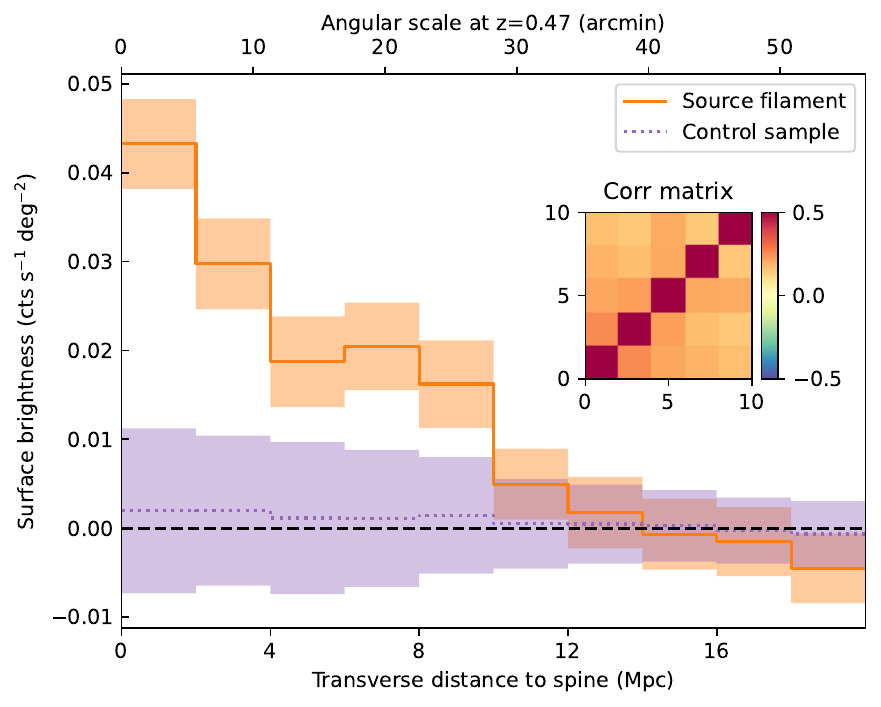}
    \caption{Stacked filament surface brightness profile in the 0.3--1.2~keV band at the observer's frame. The X-axis is the transverse distance from the filament spine. The orange and purple lines and shaded regions denote the profiles of the source and the averaged value of the random-positioned control samples, respectively. 
    The average surface brightness in the 10--20~Mpc transverse distance range is used as the local background and is subtracted from the profile to obtain the surface brightness signal of the net source profile. The shaded area of the source profile is the $1\sigma$ uncertainty estimated from a pixel-wise bootstrapping method, and that of the control samples is the $1\sigma$ scatter. The correlation matrix of the five bins in the 0--10~Mpc range is plotted in the inset figure.
    }
    \label{fig:total_prof}
\end{figure}

We applied our stacking method to the HEALPix count and exposure maps we created (as described in Sect.~\ref{sect:healpix}). Figure~\ref{fig:total_prof} shows the stacked surface brightness against the transverse distance from the filament spine in the 0.3--1.2~keV band. The profile is similar to those in \citet{Tanimura2020-xray, Tanimura2022}, exhibiting a centrally peaked signal, and the intensity gradually decreases to $\sim$10~Mpc in the transverse distance. We used the 10--20~Mpc distance region to evaluate the local background level and subtracted it from the profile to obtain the net surface brightness profile of the filaments. To evaluate the significance of detecting the source signal by accounting for the magnitude of fluctuations and systematic effects due to varying background levels across the sky, we constructed 500 control samples of filaments. Each control sample contains 7817 filaments, but we moved them to random positions in the analysis footprint and then rotated them by random angles. The mean surface brightness profile and the corresponding standard deviation of these 500 control samples are also plotted in Fig.~\ref{fig:total_prof}. The excess X-ray emission from the source filaments over the control sample (i.e.,background in this case) is significant, reaching a $9.0\sigma$ confidence level in the 0--10~Mpc bins. 

To investigate the dependence of the significance of the stacked profile on the selection of the energy band and filament lengths, we applied our analysis to two sub-bands and two distinct filament populations of different lengths and found consistent results with our analysis with the default selection (see Appendix \ref{app:profile_more} for details). The signals in the 0.3--0.7~keV and the 0.7--1.2~keV bands are significant. Meanwhile, the detected filament width of the 20--40~Mpc lengths is marginally broader than that of the 40--100~Mpc length.

Furthermore, we investigated the impact of different source mask radii on the stacked profile (see Appendix \ref{app:mask_radius} for details). If we increase the cluster mask radii by a factor of two, the resulting profile agrees with the original one within the uncertainties and the integrated signal within the 0--10~Mpc transverse distance bin is only reduced by 7\% (see the left panel of Fig. \ref{fig:prof_mask_size}). If we adopt a more conservative mask radius criteria for eRASS:4 X-ray sources, which is $S(r_\mathrm{cut})=0.05S_\mathrm{bkg}$, the resulting profile agrees with the original one within $1\sigma$ uncertainties as well and the integrated signal is enhanced by 3\% (see the right panel of Fig. \ref{fig:prof_mask_size}). Therefore, we conclude that increasing the source mask sizes has a trivial impact on the net profile normalization, which is in agreement with the inspections presented in \citet{Tanimura2022}.

\subsection{Stacking method validation}\label{sect:validation}

\begin{figure}
    \centering
    \includegraphics[width=0.45\textwidth]{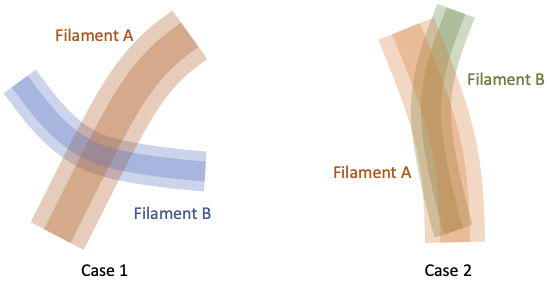}
    \caption{Sketches of two types of overlapped filaments due to the projection effects. Deep and shallow regions are the source and local background regions, respectively. In the first case (\emph{left}), the two filaments have different orientations, and the emission in the overlapping region will be accounted for by local background subtraction. In the second case (\emph{right}), the two filaments have similar orientations and will lead to overestimated stacked results, evaluated in Sect. \ref{sect:validation}.}
    \label{fig:overlapping}
\end{figure}

Filaments at different redshifts could be overlapped with each other due to projection effects. The average number of projected filaments on the pixels within 4~Mpc transverse distance for our selected filament sample is 2.5. This number increases to 5.5 and 11 for pixels within transverse distances of 10~Mpc and 20~Mpc, respectively.
There are two different cases of overlapping. In the first case, the two overlapped filaments have different orientations (left in Fig.~\ref{fig:overlapping}). When extracting the profile of filament A, the contribution from filament B in the overlapped source region will be accounted for by the filament B emission in the local background region and vice versa. In the second case, the two overlapped filaments have similar orientations (right in Fig.~\ref{fig:overlapping}), while the emission in the overlapping source region will be double counted; therefore increasing the stacked signal. Meanwhile, our stacking method assumes that systematic uncertainties from the spatial variations of multiple components, for instance, the Milky Way foreground emission and absorption,\ac{cxb} emission. In addition, the inhomogeneity of exposure is minimized by a large number of filaments randomly distributed in the footprint. 

\begin{figure}
    \centering
    \includegraphics[width=0.49\textwidth]{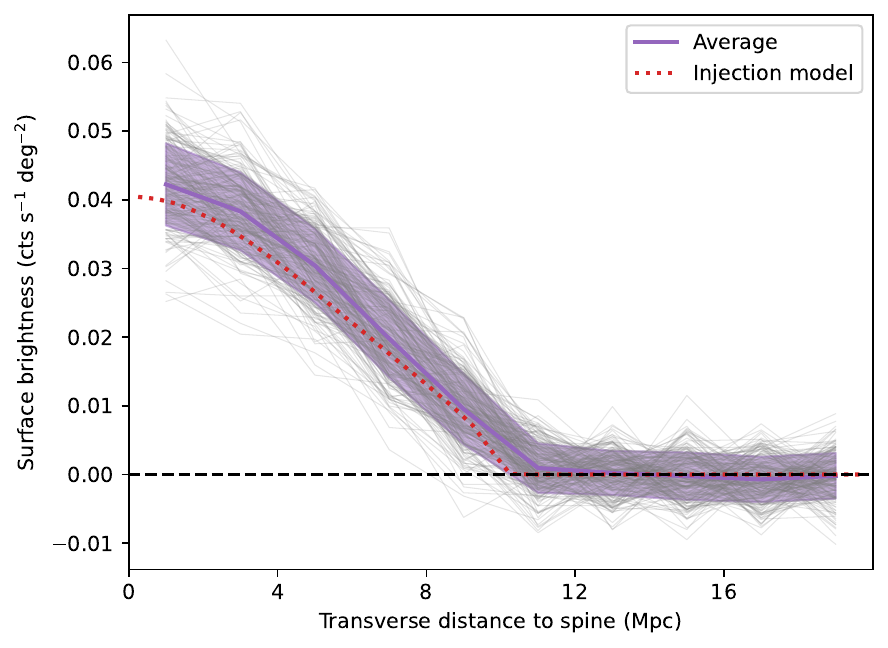}
    \caption{Results of 200 net surface brightness profiles of the simulated maps obtained by applying our stacking method (in gray). The red dashed curve is the injected surface brightness profile model. The curve in purple and the shaded region are the mean and standard deviation of the 200 profiles. Although the filament signal at the spine is only $\sim0.5\%$ of the background value, the stacking method can recover the injected profile successfully.}
    \label{fig:mock_profile}
\end{figure}

To evaluate the impact of the second case overlapping and multiple possible systematic uncertainties, we first simulated the mock count maps and applied the same surface brightness profile stacking method to the mock maps to obtain the averaged stacked result and to compare it with the injected model. The details of the mock map creation are presented in Appendix \ref{app:mock_map}. 
We simulated 200 mock count maps and applied the same profile stacking method. The stacked mock profiles from the mock maps are plotted in Fig.~\ref{fig:mock_profile}, together with the injected model profile. The averaged mock profile is in general agreement with the injected model within the scatter; whereas, it is slightly higher than the injected model by $13\%$. This $13\%$ enhanced signal could be due to the second case of overlapping and we corrected this effect when modeling the gas properties (described in the subsequent sections). The validation also suggests that although the averaged filament signal at the spine is $0.5\%$ of the total background level, our stacking method can successfully recover the emission profile. 

\section{Modeling the contamination from unresolved extragalactic sources}\label{sect:contam}
The source masks applied in this work successfully mask out all X-ray-detected point sources and galaxy clusters and groups above the eRASS:4 limit, as well as optically selected galaxy clusters and groups with $\lambda>20$. We note that X-ray sources outside these selection criteria (i.e., those below the eRASS:4 detection limit and low-richness clusters and groups) may also contribute to the stacked X-ray signal. To remove the contaminating signal, we evaluated the contribution from three types of unmasked X-ray signal associated with galaxies: gravitationally bound hot gas in galaxy or galaxy group halos, undetected low-luminosity \acp{agn}, and \acp{xrb}. The steps required to model the contribution from these sources in the stacked total signal are given below.

\begin{enumerate}
    \item We built a redshift-dependent $L_\mathrm{X}-M_*$ scaling relation of each type of the sources in a $9.5<\log(M_*/M_\odot)<12$ stellar mass range. We describe the contribution from each type of unmasked source in detail in Sect.~\ref{sect:l-m_individual}.
    \item We adopted the DESI Legacy Survey DR9 galaxy photometric redshift and stellar mass catalog \citep{Zou2019}, which is relatively complete and covers our analysis footprint. We selected galaxies in the redshift range of [0.15, 0.7] and cleaned up this catalog by removing sources with a redshift uncertainty of $>20\%$. We predicted the expected count rate in the 0.3--1.2~keV band for each galaxy using the constructed scaling relations, given the stellar mass, redshift, spectral shape of the source, and the Galactic absorption at the galaxy's position. We further corrected for the incompleteness of the observed \ac{smf} from \citet{Zou2019} catalog to a fiducial empirical \ac{smf} model from the result of UniverseMachine \citep{Behroozi2019}, which is a project characterizing the link between galaxy growth and halo growth. In the $M_*<10^{11}M_\sun$ range, where the galaxies in \citet{Zou2019} catalog are not complete, the ratio between the \ac{smf} of the selected galaxies from \citet{Zou2019} and the fiducial \acp{smf} at different stellar masses and redshifts are used to correct the predicted count rate of each galaxy. 
    \item For each selected galaxy, we added the predicted count rate to the HEALPix pixel that the galaxy is located in. Thanks to this method, we were able to create a predicted count rate map of unmasked galaxy X-ray emission in the selected redshift range. We applied the same source mask used in the profile extracting and stacking pipeline to extract the profile of unmasked signals. 
\end{enumerate}
\begin{figure*}
    \sidecaption
    \includegraphics[width=12 cm]{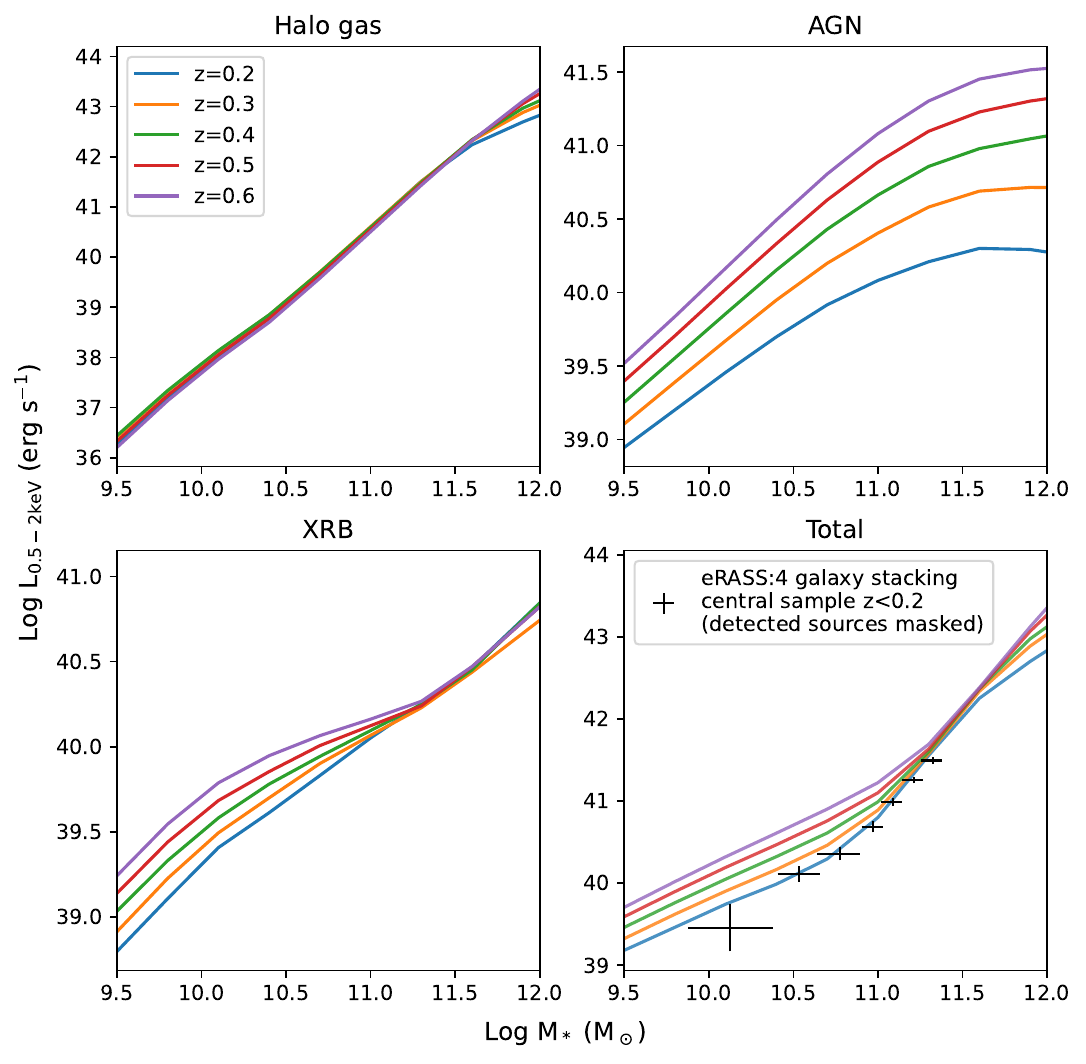}
    \caption{Rest-frame galaxy averaged 0.5--2~keV luminosity - stellar mass scaling relations for unmasked X-ray halos (\emph{top-left}), \acp{agn} (\emph{top-right}), and \acp {xrb} (\emph{bottom-left}) at different redshifts. The scaling relations of total unmasked X-ray signals associated with galaxies are plotted (\emph{bottom right})  together with the recent eRASS stacking results of $z<0.2$ central galaxy sample in black. }
    \label{fig:Lx-Ms}
\end{figure*}
\subsection{Properties of contaminating unmasked sources}\label{sect:l-m_individual}

To estimate the flux distribution of the contaminating signal, we constructed $L_\mathrm{X}-M_*$ for three types of unmasked X-ray sources. Meanwhile, we also investigated the spectral properties to be used in count rate conversion in spectral fitting. The details of the modeling for each component are given in the subsections below.

\subsubsection{Unresolved haloes}\label{sect:halo_scling}
We used the X-ray halo $L_\mathrm{X}-M_*$ relation from \ac{rass} galaxy stacking analysis \citep{Anderson2015} to predict the X-ray luminosity of bound gas in unmasked galaxy (group and cluster) halos. Because we excluded eRASS:4-detected X-ray sources and clusters detected in the optical band with a richness higher than 20 before stacking, to obtain the $L_\mathrm{X}-M_*$ relation that meets the same selection -- we use our galaxy cluster mock catalog that is constructed for \ac{erass} predictions. The catalog is based on HugeMDPL dark matter-only N-body simulation \citep{Klypin2016}, while the X-ray properties of each halo are illustrated by a phenomenological modeling approach \citep{Comparat2020} and then scaled to agree with the observed $L_\mathrm{X}-M_*$ relation from \ac{rass} galaxy stacking results of \citet{Anderson2015}. We tabulated the averaged rest-frame $L_\mathrm{0.5-2keV}-M_*$ relation for unmasked X-ray halos in different redshift bins by applying the same source selection criteria. 

The galaxy sample used in \citet{Anderson2015} was selected from an \ac{sdss}-DR7 based \ac{nyu-vagc} \citep{Blanton2005}.
The stellar mass in \ac{nyu-vagc} is calculated based on the method of \citet{Blanton2007}. It is known that the \citet{Blanton2007} stellar mass is systematically lower than many other estimates at the high stellar mass end \citep[e.g.,][]{Li2009,Bernardi2010,Moustakas2013}. Because the X-ray halo $L_\mathrm{X}-M_*$ relation has a steep slope of $\gtrsim3$, a small underestimate of the stellar mass at the high-mass end could introduce a huge difference on the estimated unmasked X-ray halo emission. Therefore, we need to properly modify the current scaling relation to match the \citet{Zou2019} stellar mass scheme. We selected a $z<0.2$ subsample of \citet{Zou2019} galaxies with available spectroscopic redshift to cross-match the \ac{nyu-vagc} catalog and obtained the discrepancy between the two stellar mass schemes. We used this stellar mass discrepancy to empirically modify the tabulated $L_\mathrm{0.5-2keV}-M_*$ relation. The modified X-ray halo $L_\mathrm{X}-M_*$ relation is plotted in the top left panel of Fig.~\ref{fig:Lx-Ms}).

To convert the rest-frame 0.5--2~keV luminosity to the 0.3--1.2~keV observed count rate, the spectral properties of the sources are needed. For this purpose, we assumed that each source can be represented by a single temperature \ac{cie} model and we estimated the temperature using the luminosity - temperature ($L_\mathrm{X}-kT$ ) scaling relation constructed using \ac{efeds} galaxy cluster and groups \citep{Bahar2022}. 

With the estimated temperature of each source in the mock catalog by using the $L_\mathrm{X}-kT$ scaling relation, we were also able to predict a synthesis spectral model of unmasked X-ray halos for spectral analysis. After applying the selection, we construct the multi-temperature \ac{dem} models for each redshift bin (left panel of Fig.~\ref{fig:spec-dem}). The spectral model on each single temperature is calculated using \ac{apec} with $Z=0.4Z_\sun$. Then, we used the redshift and length distributions of the selected filament sample to create the synthesis spectral model as:
\begin{equation}
    F(E) = \sum_i \frac{l_i\times F(E, z_i)}{D_\mathrm{L}(z_i)^2},
\end{equation}
\noindent where $l_i$ is the physical length of the $i$-th filament, $z_i$ is the redshift of the $i$-th filament, $F(E,z_i)$ is the multi-temperature differential emission measure spectrum at the redshift of $z_i$, and $D_\mathrm{L}$ is the luminosity distance as a function of redshift. The constructed model spectrum is shown in the right panel of Fig.~\ref{fig:spec-dem}.

\subsubsection{Unresolved \ac{agn}}
We calculate the galaxy averaged $L_\mathrm{X}-M_*$ relation for unresolved \acp{agn} using an \ac{agn} mock catalog based on the high-resolution Uchuu dark matter only N-body simulation \citep{Ishiyama2021}. The AGN properties are painted to dark matter halos based on an empirical method \citep[see][for further details]{Comparat2019}, where the \ac{agn} model is based on the observed \ac{agn} 2--10~keV X-ray luminosity function \citep{Aird2015}, duty cycle \citep{Georgakakis2017}, obscuration distribution \citep{Ricci2017}, spectral model \citep{Liu2017}, and parameters of \ac{agn} halo occupation distribution \citep{Comparat2023}. 

We applied a selection criteria that removes \acp{agn} with $F_\mathrm{0.5-2keV}>10^{-14}$~erg~s$^{-1}$~cm$^{-2}$, which would be detected in eRASS:4, to select sources in the mock catalog. In each redshift-stellar mass bin, we summed the rest-frame 0.5--2~keV luminosity and divide it by the total number of galaxies in the same bin to obtain the galaxy averaged unresolved \ac{agn} luminosity. In our method, a galaxy-averaged scaling relation would distribute the unmasked \ac{agn} flux to all galaxies. Therefore, we do not need to additionally predict whether a galaxy hosts an \ac{agn}. The scaling relations at different redshifts are shown in the top right panel of Fig. \ref{fig:Lx-Ms}.

We also investigated the stacked spectral properties of unresolved \acp{agn}. We adopted the same spectral model used to create the mock catalog. The detailed components in XSPEC \citep{Arnaud1996} are \texttt{plcabs} + \texttt{zgauss} + \texttt{constant} * \texttt{powerlaw} + \texttt{pexrav} * \texttt{constant}, representing a power-law model with intrinsic absorption, the 6.4~keV Fe~K$\alpha$ emission line, a soft-X-ray-excess, and a cold-reflection component, respectively. We predicted a corresponding mock spectrum based on the hard band luminosity and the intrinsic absorbing column density for each unmasked source. The stacked spectra of different bins and the median spectrum are plotted in Fig.~\ref{fig:agn-spec}. They show a strong self-similarity in the soft band. Therefore, we use the median spectrum to convert the stacked rest-frame 0.5--2~keV luminosity to 0.3--1.2~keV count rate.

\subsubsection{\ac{xrb} in galaxies}
We used an integrated X-ray luminosity model that is a function of stellar mass and \ac{sfr} to construct the averaged $L_\mathrm{X}-M_*$ relations of \ac{xrb} at different redshifts.
The model is constrained by \emph{Chandra} observations of local galaxies \citep{Lehmer2019}, and the UniverseMachine model of redshift dependent \ac{smf} and \ac{sfr} \citep{Behroozi2019}. 
The results are plotted in the bottom left panel of Fig.~\ref{fig:Lx-Ms}, where the relations show redshift evolution at $M_*<10^{11}$~$M_\odot$ (the result of the redshift evolution of star-forming rate and star-forming galaxy fraction). 

We adopted a power-law distribution with intrinsic absorption spectral model for flux conversion, with parameters of $\Gamma=1.7$ and $n_\mathrm{H}=2\times10^{21}$ cm$^{-2}$, as suggested by observations \citep{Lehmer2019}.

\subsubsection{Comparison with observed $L_\mathrm{X}-M_*$ relations}
 \citet{Zhang2024} investigated the $L_\mathrm{X}-M_*$ scaling relation of galaxies using eRASS:4 data. We compared our constructed total unresolved galactic X-ray sources and their $L_\mathrm{X}-M_*$ relations with their results from eROSITA observations. We note that the stellar mass used in \citet{Zhang2024} was adopted from \citet{Tinker2021}, which was originally estimated by \citet{Chen2012}. Similarly to the analysis in Sect. \ref{sect:halo_scling}, we scaled the observed $L_\mathrm{X}-M_*$ scaling relation in \citet{Zhang2024} to match the stellar mass scheme of \citet{Zou2019} for comparison. In the bottom right panel of Fig.~\ref{fig:Lx-Ms}, we plot the total unresolved source scaling relation model together with that of a central galaxy sample from \citet{Zhang2024}, whose median redshift is $\sim0.1$. Our constructed relation at $z=0.2$ has the same slope at different stellar masses as the observed relation and the slight luminosity difference meets the expected redshift revolution of \ac{agn} and \ac{xrb} scaling relations. 

\subsection{Fractions in stacked signals}\label{sect:unmasked-fraction}

\begin{figure}
    \centering
    \includegraphics[width=0.49\textwidth]{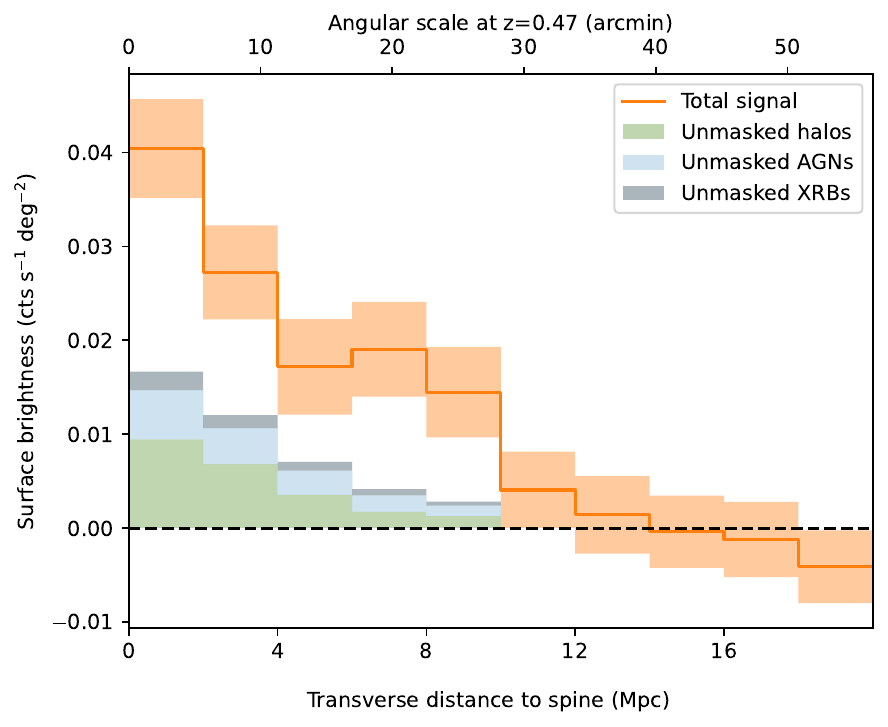}
    \caption{Contribution of unmasked X-ray halos (green), \acp{agn} (light blue), and \acp{xrb} (grey) to the total stacked signal (orange). To find the net \ac{whim} signal, the contribution from these contaminating signals is modeled out in the analysis. The signal from WHIM is $\sim60\%$ based on the unmasked signal estimation. }
    \label{fig:contam_prof}
\end{figure}

Next, we calculated the integrated signal of unmasked halos, \acp{agn}, and \acp{xrb} in the 0--10~Mpc transverse distance range to account for the contamination in our imaging analysis. The total contribution is $37\%$, with each of $20\%$, $12\%$, and $5\%$ for unmasked halos, \acp{agn}, and \acp{xrb}, respectively (also see Fig.~\ref{fig:contam_prof}). 
We further divided the stellar mass range into two subranges ($9.5<\log M_*(M_\sun)<11$ and $11<\log M_*(M_\sun)<12$) and repeated the same analysis. The results of the fractions are listed in Table~\ref{tab:fraction_unmasked}. In the high stellar mass bin, where the unmasked galaxies are mostly central galaxies of group-size halos, the unmasked signal is dominated by halos gas; whereas in the low stellar mass bin, unmasked \ac{agn}s and \ac{xrb}s demonstrate more of a contribution. We used a bootstrapping method to estimate the statistical uncertainties of the fraction. In short, we randomly select galaxies with replacements to construct bootstrapping samples and calculated the standard deviations of the fractions estimated using the bootstrapping samples as the statistical uncertainties. The relative uncertainties are $6.9\%$, $1.8\%$, and $1.4\%$ for unmasked halos, \acp{agn}, and \ac{xrb}, respectively. Given that the unmasked halos contribute about half of the unmasked signal, the averaged statistical uncertainty is below $5\%$. 

Because the uncertainty on photometric redshifts that we used in this analysis is not negligible and is a function of magnitude (it can reach up to a few tens of percent for faint galaxies \citealt{Zou2019}), it has a potential impact on the recovered unresolved contamination signal. The redshift range of the filament sample we select is $0.2<z<0.6$, while the range for the galaxies we used to model unmasked source contribution is $0.15<z<0.7$. This means that for a galaxy whose true redshift $z_\mathrm{true}$ is within the filament redshift range, but the measured redshift $z_\mathrm{phot}$ is outside the redshift range of the selected galaxies, it may not be taken into account in the unmasked source profile. On the contrary, for a galaxy with $z_\mathrm{true}$ outside of the filament redshift range but $z_\mathrm{phot}$ within that range, because it is not spatially correlated with the filament we selected, it may not contribute to the unmasked source profile either. Therefore, we note that the fraction of unmasked source signals could be underestimated in our analysis due to the galaxy photo-z uncertainties. 

\begin{figure}
    \centering
    \includegraphics[width=0.48\textwidth]{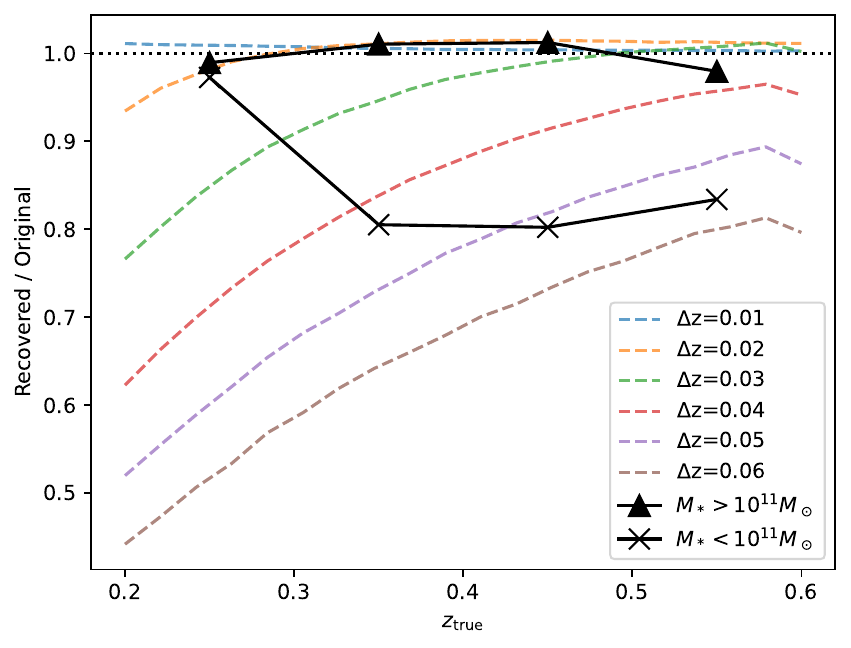}
    \caption{Fraction of recovered flux of unresolved sources as a function of true redshift (x-axis) and normalized redshift uncertainty (dashed lines in color). The triangle and crossed lines are the fractions for our selected galaxies in the $11<\log M_*(M_\sun)<12$ and $9.5<\log M_*(M_\sun)<11$ stellar mass bins, respectively. The impact of photometric redshift uncertainty is only a few percent for high stellar mass galaxies that contribute most of the unmasked signals.} 
    \label{fig:recover_fl}
\end{figure}

We use a Monte Carlo method to calculate the recovered flux fraction of galaxies at a given redshift with given redshift uncertainties. For creating the randomized redshifts, we follow the definition of normalized redshift error used by \citet{Zou2019}, $\Delta z_\mathrm{norm}\equiv(z_\mathrm{phot}-z_\mathrm{spec})/(1+z_\mathrm{spec})$ and assume that $z_\mathrm{spec}=z_\mathrm{true}$.  
The ratio between the recovered flux and the true flux of a galaxy is 
\begin{equation}
    f=\frac{D_\mathrm{L}^2(z_\mathrm{true})}{n_\mathrm{MC}}\sum_i{\frac{w_i}{D_\mathrm{L}^2(z_i)}}, 
\end{equation}
where $n_\mathrm{MC}$ is the number of realizations, and $z_i$ is the $i$~th randomized redshift value, while the value of $w_i$ is 1 if $0.15<z_i<0.7$ and it is 0 otherwise. 
The results are shown as dashed lines in Fig.~\ref{fig:recover_fl}. Based on the $90\%$ percentile of the $r$-band magnitude distribution of the galaxies we selected and the $r-\Delta z$ relation shown in Fig. 5 of \citet{Zou2019}, we were able to predict the recovered flux fractions for unmasked galaxies in the $9.5<\log M_*(M_\sun)<11$ and $11<\log M_*(M_\sun)<12$ stellar mass ranges, which are shown in the crossed and triangle curve in Fig.~\ref{fig:recover_fl}, respectively. The recovered fraction of galaxies in the high stellar bin is around one due to the high quality of photometric redshift of bright central galaxies. This fraction is $\sim80\%$ in the low stellar mass bin when $z>0.3$. It means that even if we take the impact of photo-z uncertainty into account, the total contribution from unresolved sources can be as high as $\sim40\%$. 

The $\sim40\%$ fraction of the unmasked source contribution suggests that the remaining $\sim60\%$ of the stacked signal is from the \ac{whim} in the filaments, reaching a $5.4\sigma$ detection significance. 
This fraction relies on the galaxy X-ray contamination modeling, which is based on our current knowledge of different X-ray emitters in galaxy systems at different scales. A  better understanding of all the different types of X-ray sources in the Universe will offer more accurate results of the contamination fraction in the future.

\begin{table}[]
\caption{Fraction of X-ray signal from unmasked sources in different stellar mass bins.}\label{tab:fraction_unmasked}
    \centering
    \begin{tabular}{cccc}
        \hline\hline
        $\log (M_*/M_\sun)$ & Halo & \ac{agn} & \ac{xrb} \\ 
        \hline
        9.5--12 & 0.20 & 0.12 & 0.05 \\
        9.5--11 & 0.01 & 0.09 & 0.04 \\
        11--12 & 0.19 & 0.03 & 0.01 \\
        \hline
\end{tabular}
\end{table}
\section{Spectral stacking}\label{sect:spec}
In this section, we describe the X-ray spectral analysis of the filaments. Due to the faint nature of the cosmic filaments, to increase the signal-to-noise ratio (S/N) in the data, we stacked the spectra extracted from individual filaments by following the recipe provided in \citet{Bulbul2014}. The method successfully produces the stacked rest-frame spectrum of filaments at different redshifts. Through this analysis, we study the spectral properties and constrain the temperature of the detected \ac{whim} using the stacked filament spectra.

\subsection{Stacking processes}
For each filament, we first merged the event files of TM8 and shifted the pulse-invariant of each event by a factor of $1+z$. Then, we used \texttt{srctool} to extract the source and local background spectra and the vignetting corrected source \ac{arf}. The source and local background extraction regions are defined as 0--10~Mpc and 10--20~Mpc transverse distance to the filament spine, whose region masks for spectral extraction are created by constructing images using and excluding sources in Table~\ref{tab:source_mask}. For each \ac{arf}, we first corrected it by multiplying the foreground absorption curve given the averaged Galactic $n_\mathrm{H}$ value \citep{HI4PI2016} in the extraction region and the TBABS absorption model \citep{Whilms2000}. We then shifted it by a factor of $1+z$ to the source rest frame. The \ac{rmf} was taken from the calibration database and then shifted to the source rest frame by adopting a two-dimensional (2D) interpolation strategy on both axes of induced photon energy and resulting spectral channels \citep[see][for further details]{Bulbul2014}.

We used tools \texttt{addspec}, \texttt{addarf}, and \texttt{addrmf} in FTOOLS\footnote{\url{http://heasarc.gsfc.nasa.gov/ftools}} \citep{Heasarc2014} to stack the spectra, \acp{arf}, and RMFs, respectively. 
We note that some keywords in the headers of \texttt{srctool} products have different definitions from those in other X-ray satellite data products\footnote{See \url{https://erosita.mpe.mpg.de/edr/DataAnalysis/srctool_doc.html\#Output_files} for the details of the \texttt{srctool} products.}. In short, the EXPOSURE value of a spectrum $t_\mathrm{EXP}$ is the total time of an extraction region on a detector, instead of the averaged exposure time in the region $t_\mathrm{ave}$; BACKSCAL value $\Omega_\mathrm{BS}$ is the averaged intersection of the extraction region and the detector area; the geometric solid angle of the extraction region $\Omega_\mathrm{REG}$ has a keyword of REGAREA, along with the ratio $\Omega_\mathrm{REG}/\Omega_\mathrm{BS}=t_\mathrm{EXP}/t_\mathrm{ave}$. Furthermore, the normalization of an \ac{arf} curve is scaled by a factor of $t_\mathrm{ave}/t_\mathrm{EXP}$. Therefore, we need to add up all individual source spectra to obtain a stacked source spectrum, as follows: 

\begin{equation}
    N_\mathrm{stacked,src}(E)=\sum_i N_{\mathrm{src},i}(E).
\end{equation}
\noindent We then scale each local background spectrum to match the extraction solid angle of the corresponding source spectrum to get the stacked background spectrum,
\begin{equation}
    N_\mathrm{stacked,bkg}(E)=\sum_i \frac{\Omega_{\mathrm{REG,src},i}}{\Omega_{\mathrm{REG,bkg},i}}\times N_{\mathrm{bkg},i}(E).
\end{equation}
For the final stacked source spectrum, we assign the exposure time as a solid angle weighted averaged exposure $\overline{t}_\mathrm{ave}=\left(\sum_i \Omega_{\mathrm{REG},i}\times t_{\mathrm{EXP},i}\right)/\sum_i \Omega_{\mathrm{REG},i}$.
To match this exposure, the stacked \ac{arf} is defined as 
\begin{equation}
    A_\mathrm{stacked}(E)=\frac{1}{\overline{t}_\mathrm{ave}}\sum_i t_{\mathrm{EXP},i}\Omega_{\mathrm{REG},i}A_i(E),
\end{equation}
\noindent which takes each extraction solid angle into account. The modeled source flux using the exposure and stacked \ac{arf} is scaled per $\deg^2$. Similarly, the stacked RMF is weighted by $\Omega_\mathrm{REG}\times t_\mathrm{ave}$, which is 

\begin{equation}
    \boldsymbol{\mathcal{R}}_\mathrm{stacked}=\frac{1}{\sum_i\Omega_{\mathrm{BS},i} t_{\mathrm{EXP},i}}\times \sum_i \Omega_{\mathrm{BS},i} t_{\mathrm{EXP},i} \boldsymbol{\mathcal{R}}_i.
\end{equation}

The blueshifted response files at the rest frame corresponding to each filament are later stacked using FTOOLS \texttt{addarf} and \texttt{addrmf} to obtain one final stacked response file. These files are used to model the stacked spectrum described in the section below.

One clear advantage of the spectral stacking method in the rest frame is reducing the effect on the instrumental and background features on the final net background-subtracted source spectrum as described in \citep{Bulbul2014}. The energies of the instrumental and celestial X-ray foreground and background features, frozen in the detector frame, are smoothed over the redshift range covered by the sample, reducing their overall effect on the net spectrum. The spectral stacking method is well suited for obtaining a spectrum of faint sources, such as cosmic filaments.

\begin{figure*}
    \centering
    \includegraphics[width=0.48\textwidth]{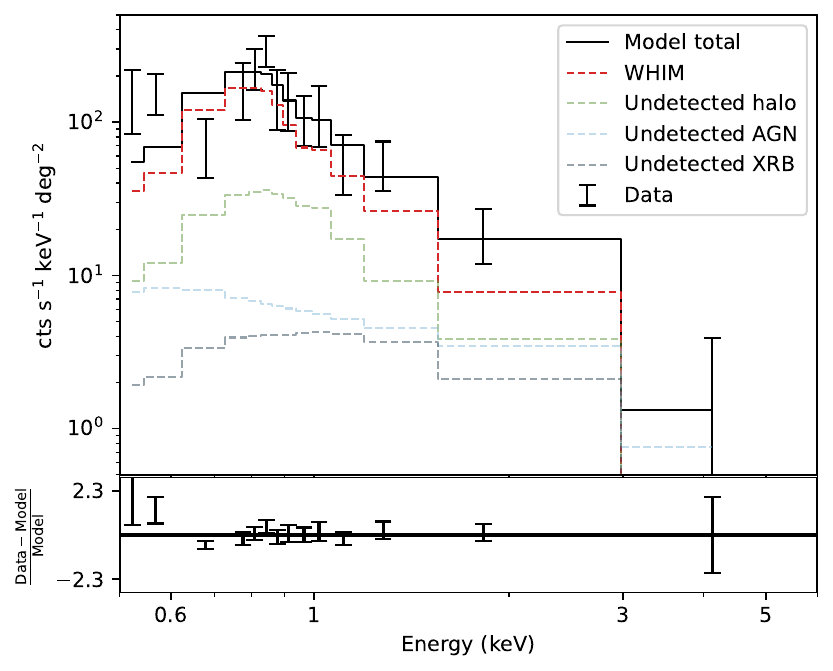}
    \includegraphics[width=0.48\textwidth]{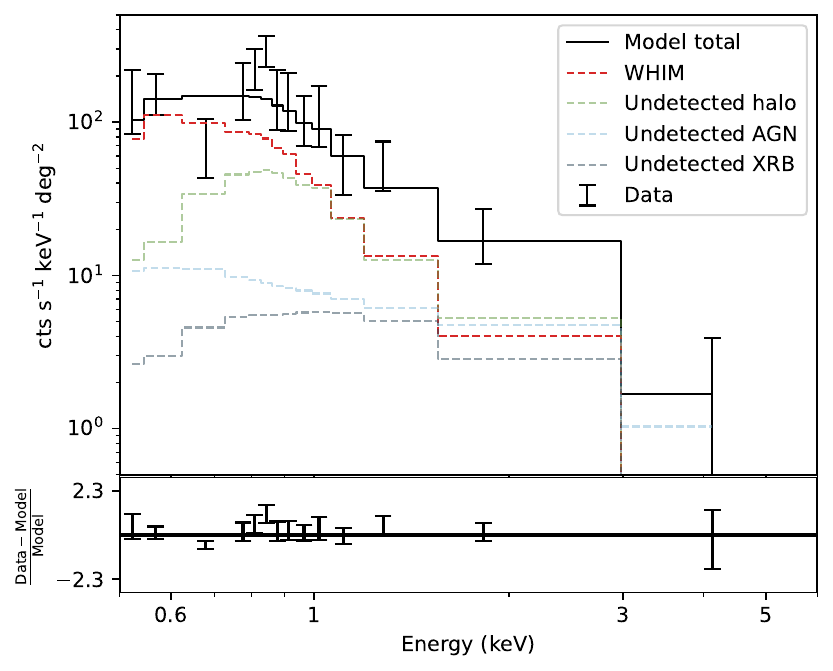}
    \caption{Rest frame stacked source spectrum of the 0--10~Mpc region from the filament spine in the transverse direction after the subtraction of the local background extracted in the 10--20~Mpc region surrounding the filament. For illustration, the spectrum is binned to a minimum signal-to-noise ratio of 2.5. The \emph{left} panel shows that the combination of a single temperature \ac{cie} (red dashed line), and the unmasked sources models (green, blue, and grey) can well fit most of the spectrum, except the significant excess at $E<0.6$~keV. The best-fit temperature of the WHIM gas is $0.58\pm0.10$~keV. 
    The \emph{right} panel shows a fitting by a log-normal temperature distribution with $\log(\mu/\mathrm{K})=5.9$ and $\sigma=0.45$ dex. The 0.5--0.6~keV energy range can be well fitted by this model, though there is a likely deficit at $\sim0.7$~keV. }
    \label{fig:spec}
\end{figure*}
\subsection{Stacked spectrum and spectral fitting}\label{sect:spec_fit}
We stacked the 0--10~Mpc transverse distance region spectra as the source spectrum and the 10--20~Mpc transverse distance region spectra as the local background region spectrum. The stacked background spectrum is smoothed using a Wiener filter with a kernel size of five channels for direct subtraction. Additionally, we compare the 6--9~keV count rate of the two spectra and store the ratio as the \textsc{backscal} in the header for further correction. The stacked net source spectrum is plotted in Fig.~\ref{fig:spec}.

We used XSPEC version 12.13.0 and the  BXA\footnote{\url{https://github.com/JohannesBuchner/BXA}} package \citep{Buchner2014} for our spectral analysis. The spectrum was first optimally binned \citep{Kaastra2016} using the tool \texttt{ftgrouppha} and then loaded into XSPEC. We used the 0.5--6.0~keV range for analysis and adopted C-stat \citep{Kaastra2017} to derive the fitting likelihood. The BXA package uses the nested sampling algorithm UltraNest \citep{Buchner2016,Buchner2019,Buchner2021} for Bayesian inference.

We fit the spectrum using four components that account for the detected \ac{whim}, unmasked halo gas, unmasked \ac{agn}, and unmasked \ac{xrb}. The spectral model of the three unresolved sources is described in Sect.~\ref{sect:l-m_individual}. We calculated the expected normalization ratios among the three components based on the modeled unresolved source fractions in Sect.~\ref{sect:unmasked-fraction} and coupled the ratios during fitting. To model the \ac{whim} component, we used a \ac{1t} \ac{apec} model with fixed abundance $Z=0.2Z_\sun$. This value is based on the reported $\lesssim0.3Z_\odot$ abundance in cluster outskirts by both observations and numerical simulations \citep{Bulbul2016,Mernier2017,Biffi2017} and $>0.1Z_\odot$ in $z<1$ filaments from numerical simulations \citep{Martizzi2019}. This model combination can well fit the $E>0.7$~keV spectrum range with a best-fit parameter $kT=0.58\pm0.10$~keV and a fitting statistics $c_\mathrm{stat}/\mathrm{d.o.f.}=105/111$. We tried to fix the metalicity to $0.1Z_\sun$ and $0.3Z_\sun$ to test the impact of the metalicity assumption and found there is no impact on the best-fit temperature. In addition, we also fitted a stacked 0--4 Mpc transverse distance region spectrum and obtained a similar $0.56\pm0.09$~keV temperature. 

Recently, \citet{Vijayan2022} demonstrated that fitting a multi-temperature nature spectrum using a \ac{1t} model will lead to an overestimation in the temperature. It explains that our best-fit temperature using a \ac{1t} model does not agree with the majority of the gas phases around $10^6$~K predicted by simulations \citep[e.g.,][]{Martizzi2019,Galarraga-Espinosa2021}. In our stacked spectrum, we observe an excess feature in 0.5--0.6~keV, which agrees with the energy range of $\sim0.56$~keV O~VII lines emitted by cooler phases of gas around $10^{6.3}$ K ($\sim0.17$~keV), suggesting the presence of cooler phases. The gas in these temperature phases only contributes to lines below 0.7~keV, while the fitting using a \ac{1t} model tends to maximize the likelihood around the peak of the observed spectrum, which is in the 0.8--0.9~keV energy range, dominated by several Fe~XVII lines emitted by $10^{6.8}$~K gas.

We also tried to fit the spectrum using a multi-temperature model. We adopted a log-normal temperature distribution \ac{dem} model, with a peak at $\log(T/\mathrm{K})=5.9$ and $1\sigma$ width of 0.45~dex (as  suggested by our analysis of numerical simulations; see more in Sect. \ref{sect:simulation}). We used a simplified assumption of \ac{cie} condition and $0.2Z_\sun$ abundance of the \ac{whim}. The log-normal \ac{dem} model can better fit the stacked spectrum in the 0.5--0.6~keV energy range (see the bottom panel of Fig.~\ref{fig:spec}), though this fitting shows the feature of a likely deficit around 0.7~keV. This could be because the true temperature distribution is more complex than the log-normal distribution we use here, which creates discrepancies in some energy ranges.

\section{Detected phases of WHIM}\label{sect:phase}
\subsection{Surface brightness profile modeling}\label{sect:prof_fitting}
\begin{table*}[]
\caption{Parameters and corresponding priors to constrain the density profile of the observed phase \ac{whim}.}
\label{tab:prior-prof}
    \centering
    \begin{tabular}{llllll}
    \hline\hline
    Parameter & Unit& Description & Range & Prior \\
    \hline
    $\log\Delta_\mathrm{b,0}$ && Central baryon density contrast in the logarithmic scale & [0.5,3.5] & $\mathcal{N}(1.5, 0.5)$\\
    $\log f_\mathrm{vol,0}$ & & Central volume fraction in the logarithmic scale & [-3, -1] & $\mathcal{N}(-2,0.3)$\\
    $\log r_\mathrm{c,n}$ & Mpc & Core radius of the gas density profile in the logarithmic scale & [-1, 0.8] & $\mathcal{N}(0.3,0.2)$\\
    $\alpha_\mathrm{n}$ & & Slope of the density profile in the outskirts & [-3,0]& $\mathcal{N}(-1,0.3)$\\
    $\alpha_\mathrm{fvol}$ & & Slope of the volume fraction profile in the outskirts & [-3,0]& $\mathcal{N}(-1,0.3)$\\
    $\log\sigma_\mathrm{sm}$ & Mpc& Gaussian smoothing kernel width in logarithmic scale & [0,0.8] & $\mathcal{N}(0.3,0.2)$\\
    \hline
    \end{tabular}    
\end{table*}

With the best-fit temperature obtained in the previous section, we fit the density profiles of the gas in cosmic filaments. In this section, we use $r$ to denote the physical radius of gas cells or particles to the filament spine in the 3D space. Based on results from numerical simulations \citep{Cui2019,Galarraga-Espinosa2021,Tuominen2021,Zhu2021}, the gas in filaments has a multi-phase nature, ranging from $10^3$ to $10^7$ K in temperature and from $10^{-6}$ to $10^{-1}$~cm$^{-3}$ in hydrogen density. Therefore, we introduced the parameter, $f_\mathrm{vol}$, to represent the fraction of the volume of the detected gas to the total filament volume. The volume-averaged emissivity of the detected phase gas at a certain radius is 

\begin{equation}
        \bar{\epsilon}_\mathrm{det}(r) = 1.20\times n_\mathrm{H}(r)^2f_\mathrm{vol}(r),
\end{equation}
\noindent where $n_\mathrm{H}$ is the hydrogen density, 1.20 is the $n_\mathrm{e}/n_\mathrm{H}$ ratio for the plasma at the detected temperature. We used a King-type profile \citep{King1962}, which is widely used to fit central flattened profiles, to model the radial change of both $n_\mathrm{H}$ and $f_\mathrm{vol}$, 
\begin{align}
    n_\mathrm{H}(r) = &n_\mathrm{H,0}\left[1 + (r/r_\mathrm{c,n})^2\right]^{\alpha_\mathrm{n}}; \\
    f_\mathrm{vol}(r) = &f_\mathrm{vol,0}\left[1 + (r/r_\mathrm{c,fvol})^2\right]^{\alpha_\mathrm{fvol}},
\end{align}
\noindent where $n_\mathrm{H,0}$ ($f_\mathrm{vol,0}$), $r_\mathrm{c,n}$ ($r_\mathrm{c,fvol}$), and $\alpha_\mathrm{n}$ ($\alpha_\mathrm{fvol}$) are the central normalization, core radius, and slope in the outskirts for $n_\mathrm{H}$ (or $f_\mathrm{vol}$), respectively. Because the radial dependencies of $n_\mathrm{H}$ and $f_\mathrm{vol}$ have a strong degeneracy, we assume that they have the same core radii, namely, $r_\mathrm{c,n}=r_\mathrm{c,fvol}$. 

In addition, we must make several assumptions. First, we assume a constant baryon density contrast $\Delta_\mathrm{b,0}$ at the filament spine for filaments at all redshifts. In this case, 
\begin{align}
    n_\mathrm{H,0}(z) = \frac{\Delta_\mathrm{b,0} \rho_\mathrm{crit} \Omega_\mathrm{b}}{1.43\times m_\mathrm{H}} \times \left(1+z\right)^3,
\end{align}
\noindent where $\rho_\mathrm{crit}$ is the critical density at $z=0$, the factor 1.43 is the ratio between total mass density and hydrogen mass density, $m_\mathrm{H}$ is the hydrogen mass, and the redshift evolution comes from the expansion of the physical scale. 
Second, there is an outer boundary of the filaments $r_\mathrm{out}$, beyond which the gas density of our detected phase is zero. In this work, we have adopted a value of 10~Mpc. Third, the observed profile is smoothed and broadened due to the position uncertainty of the skeleton constructed from incomplete galaxy samples (e.g., Fig.~11 of \citealt{Galarraga-Espinosa2021}). Including these assumptions, the projected surface brightness at a given projected radial distance, $r_\mathrm{proj}$, is then

\begin{equation}
    S_\mathrm{X,model,conv}(r_\mathrm{proj},z) = G(\sigma_\mathrm{sm}) * S_\mathrm{X,model}(r_\mathrm{proj},z),
\end{equation}
\noindent where $G(\sigma_\mathrm{sm})$ is a Gaussian convolving kernel that accounts for the smoothing effect of the observed filament width. 
The surface brightness is 
\begin{align}
    S_\mathrm{X,model}(r_\mathrm{proj},z)=& A_\mathrm{CF}^{-1}K_\mathrm{ph}(1+z)^{-3}\nonumber \\ 
    & \times\int_{-l_\mathrm{out}}^{l_\mathrm{out}}\bar{\epsilon}_\mathrm{det}\left(\sqrt{l^2+r_\mathrm{proj}^2},z\right)\Lambda \mathrm{d}l ,
\end{align}
\noindent where $l$ is depth along the line of sight; $K_\mathrm{ph}$ is the K-correction of photon spectrum given the spectral shape, redshift, and energy band; the term $(1+z)^{-3}$ is the photon dimming factor; $\Lambda$ is the cooling function calculated using the \ac{apec} model given the temperature, the assumption of $0.2Z_\sun$ abundance and the energy band; $A_\mathrm{CF}$ is the conversion factor from photon flux to count rate, which is calculated given the on-axis \ac{arf}, observation band, observed photon spectrum, and Galactic absorption curve.

When computing the fitting statistics, we constructed 7817 individual model surface brightness count rate profiles based on the redshift of each filament. We use the solid angle of each filament as the weight to obtain an averaged surface brightness profile as

\begin{equation}
    S_\mathrm{X,model,stack} = \frac{1}{N}\sum_i^{N} \Omega_{i} \times S_{\mathrm{X,model,conv},i}.
\end{equation}

\noindent The fitting statistics is defined as 
\begin{equation}
\chi^2= {\boldsymbol{S}_\mathrm{X,diff}}^\intercal\boldsymbol{\mathcal{C}}^{-1}\boldsymbol{S}_\mathrm{X,diff},
\end{equation}
\noindent where $\boldsymbol{S}_\mathrm{X,diff}=\boldsymbol{S}_\mathrm{X,net,corr}-\boldsymbol{S}_\mathrm{X,unmasked,net}-\boldsymbol{S}_\mathrm{X,model,stack}\times1.13$, is the difference between the model profile and observed net profile with the unmasked sources contamination subtracted. The factor of 1.13 takes into account the possible enhancement of the signal due to the overlapping filaments (see Sect. \ref{sect:validation}). The $\boldsymbol{\mathcal{C}}$ is the covariance matrix estimated from the bootstrapping method.

We used a Bayesian inference approach to obtain posterior distributions of the six parameters: $\log\Delta_\mathrm{b,0}$, $\log f_\mathrm{vol,0}$, $\log r_\mathrm{c,n}$, $\alpha_\mathrm{n}$, $\alpha_\mathrm{fvol}$, and $\log\sigma_\mathrm{sm}$, as listed in Table \ref{tab:prior-prof}. 
In addition to the fitting statistics $\chi^2$, the likelihood function includes an additional constraint from \ac{sz} stacking results \citep{Tanimura2020-sz}, which is necessary to break up the degeneracy between $\Delta_\mathrm{b,0}$ and $f_\mathrm{vol,0}$. The predicted Compton-$y$ strength is 

\begin{equation}\label{eq:y}
    y_\mathrm{pd}(r_\mathrm{proj}) =\left[ \frac{\sigma_\mathrm{T}}{m_\mathrm{e}c^2}\int_{-l_\mathrm{out}}^{l_\mathrm{out}} kT n_\mathrm{e}\left(\sqrt{l^2+r_\mathrm{proj}^2}\right) f_\mathrm{vol}\left(\sqrt{l^2+r_\mathrm{proj}^2}\right) \mathrm{d} l\right],
\end{equation}

\noindent where $\sigma_\mathrm{T}$ is the Thomson scattering cross section, $m_\mathrm{e}$ is the electron mass, and $c$ is the speed of light. The profiles of \citet{Tanimura2020-sz} show similar broadening effects as our stacked X-ray profile. Therefore, we also convolved the predicted Compton-$y$ profile with the same Gaussian kernel to create a smoothed Compton-$y$ profile as
\begin{equation}
    y_\mathrm{pd,conv} = G(\sigma_\mathrm{sm}) * y_\mathrm{pd}.
\end{equation}

Given that the phases of the \ac{whim} we detect only contribute a fraction of the total stacked \ac{sz} signal, we tentatively set the upper limit of the $\log y_\mathrm{pd,conv}(0)$ to be -7.5, which is approximately the stacked SZ signal at filament spine reported by \citep{Tanimura2020-sz}. We used a likelihood term $P(\log y_\mathrm{pd,conv}(0))=\mathcal{N}(\mu_\mathrm{logy},\sigma_\mathrm{logy})$ as the addition constraint. We do not know how much of the total stacked Compton-$y$ is contributed by the X-ray-emitting gas. Nevertheless, we adopt $\mu_\mathrm{logy}=-8.3$ and $\sigma_\mathrm{logy}=0.5$ in our analysis. 
The combined likelihood function is

\begin{equation}
\ln \mathcal{L} = 
\begin{cases}
    -\frac{1}{2}\chi^2 - \frac{1}{2}\frac{\left[y_\mathrm{pd,conv}(0) - \mu_\mathrm{logy}\right]^2}{\sigma_\mathrm{logy}^2} & \text{when\ } \log y_\mathrm{pd,conv}(0) <= -7.5,\\
    -\infty & \text{elsewhere.}
\end{cases}
\end{equation}

We used a \ac{mcmc} method and the python package Zeus\footnote{\url{https://zeus-mcmc.readthedocs.io/en/latest/}} \citep{Karamanis2021} to perform Bayesian inference. 
The marginalized posterior distributions shows that the  $\log\Delta_\mathrm{b,0}=2.28\pm0.22$ and $\log f_\mathrm{vol,0}=-1.91\pm0.28$. By using all \ac{mcmc} chains for calculating the averaged baryon density contrast, we obtained $\log\Bar{\Delta}_\mathrm{b}=1.88\pm0.18$, which corresponds to a hydrogen density $4.4^{+2.2}_{-1.5}\times10^{-5}$~cm$^{-3}$ at the median redshift of the selected filament sample $z=0.47$. We note that we assumed a $Z=0.2Z_\sun$ metalicity for this inference. If the \ac{whim} has a metalicity of $0.1Z_\sun$ (and $0.05Z_\sun$), at the temperature of 0.6~keV, the radiative cooling function in our analysis band is a factor of 1.5 (or 2.5) smaller than that of the $0.2Z_\sun$ assumption, which would introduce a 0.09 (and 0.2) dex increase in the inferred $\log\Delta_\mathrm{b,0}$.

\begin{figure*}
    \centering
        \includegraphics[width=0.99\textwidth]{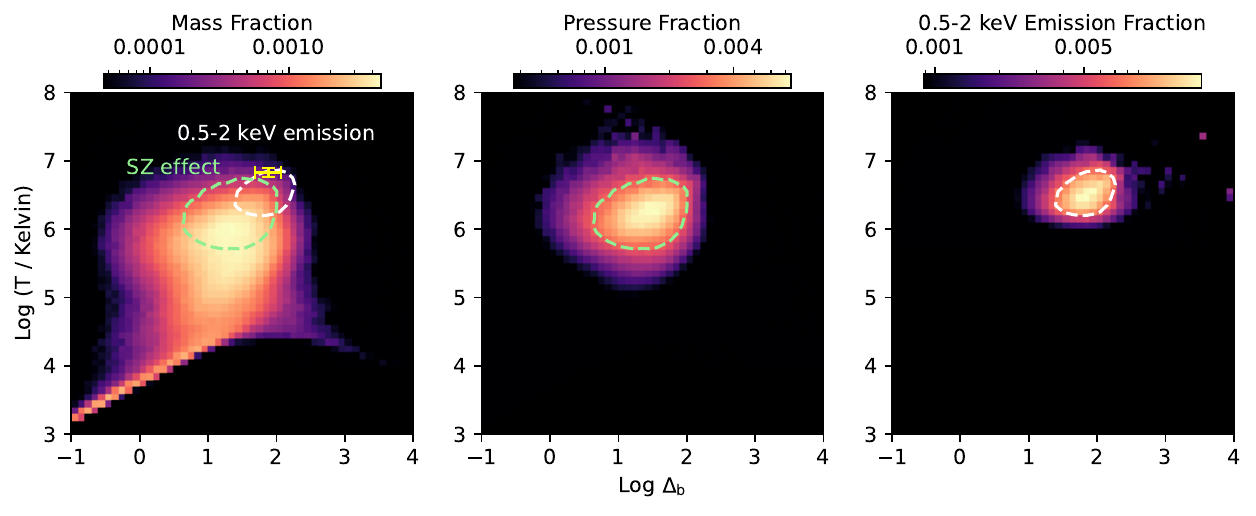} 
    \caption{IllustrisTNG phase diagrams of the filament gas at $z=0.48$ weighted by mass (\emph{left}), electron pressure (\emph{middle}), and 0.5--2~keV photon luminosity folded by eROSITA response (\emph{right}). Green and white contours are 50\% enclosed phases in electron temperature and X-ray emission, respectively. The best-fit temperature using a \ac{1t} model and baryon density contrast of this work are shown in yellow with error bars.}
    \label{fig:phase-diagram}
\end{figure*}
\subsection{X-ray emitting phases in numerical simulations}\label{sect:simulation}

To compare the measured physical quantities with those in numerical simulations, we used the public data \citep{Nelson2019} from IllustrisTNG simulations \citep{Nelson2018,Pillepich2018,Naiman2018,Springel2018,Marinacci2018}. We selected the $z=0.48$ snapshot of the TNG300-1 box and aimed to create phase diagrams of the filament gas. We used the DisPerSE algorithm to construct the filaments in that snapshot using $9<\log(M_*/M_\sun)<12$ galaxies \citep{Nelson2019}. The detailed procedure follows \citet{Galarraga-Espinosa2021} and \citet{Malavasi2022}. The calculations of the distances of each gas cell to the nearest filament, $d_\mathrm{fil}$, to the nearest maxima or bifurcation critical point, $d_\mathrm{CP}$, and the nearest critical point but along the direction of filament, $d_\mathrm{skel}$, follow the definition in \citet{Malavasi2022}. 

To reduce the computational time, we randomly selected 1/500 of the total gas cells, which is enough to represent the phase diagram \citep{Galarraga-Espinosa2021}. We selected gas cells with $d_\mathrm{fil}<1$~Mpc and $d_\mathrm{CP}>2$~Mpc. Then, we removed cells in halos or within a 25~Mpc range to simulation box boundaries. 
The temperature of a gas cell is calculated by using the quantities \texttt{InternalEnergy} and \texttt{ElectronAbundance} based on the standard recipe\footnote{\url{https://www.tng-project.org/data/docs/faq/\#gen6}}. The baryon density contrast of a cell is converted from the density by comparing it with the mean baryon density at the redshift of 0.48. 

We plot the mass-weighted phase diagrams of filament gas in the left panel of Fig.~\ref{fig:phase-diagram}, where the majority of the gas is in $10^{4.5}-10^{6.5}$~K temperature and $1-100$ baryon density contrast phases. We note that the presence of hot phases with $T>10^7$~K is not as prominent as those in the $z=0$ snapshot \citep{Galarraga-Espinosa2021}. It could be due to the redshift evolution of filament gas, which is being continuously heated during structure formation \citep{Cui2019,Martizzi2019}.

To investigate the traceable populations of the filament gas, we created two phase diagrams weighted by electron pressure and $0.5-2$~keV photon luminosity folded by eROSITA response, respectively (middle and right panels in Fig.~\ref{fig:phase-diagram}). The former one represents the traceable population of \ac{sz} observations \citep[e.g.][]{deGraaff2019,Tanimura2019,Tanimura2020-sz}, while the latter one represents the traceable population of soft band X-ray emission studies. The two diagrams show that the traceable phases of \ac{sz} signal and soft band X-ray emission differ. \ac{sz} observations trace $10^{5.5}-10^{6.8}$~K temperature and a $\Delta_\mathrm{b}=10^{0.3}-10^2$ density phases \ac{whim} in filaments.On the other hand, X-ray emission in the 0.5--2~keV band traces $10^{6.2}-10^{6.9}$~K temperature and $\Delta_\mathrm{b}=10^{1.2}-10^{2.2}$ density phases \ac{whim}. It also shows that the X-ray emitting population only contributes a fraction of the total \ac{sz} signal, which agrees with the assumption we used in Sect.~\ref{sect:prof_fitting}. 
Our averaged density contrast measurement agrees with the TNG numerical simulations. However, the best-fit temperature using a single component model is close to the upper boundary of the contour of X-ray emitting \ac{whim} (see the yellow data point with error bar in the left panel of Fig.\ref{fig:phase-diagram}). This discrepancy has been addressed in Sect. \ref{sect:spec_fit}, i.e., the best-fit temperature value using a \ac{1t} model will be biased to the high end of the temperature distribution when fitting a spectrum with multi-temperature nature. 

\subsection{Impact of photoionization}

It is known that the \ac{whim} is also partially ionized by the cosmic \ac{uvb}, which alters the ionization balance and exhibits different spectral features from the \ac{cie} plasma \citep[e.g.,][]{Nicastro1999, Krongold2003,Khabibullin2019,Churazov2023}. In this section, we describe how we combined the \ac{whim} phase distributions in numerical simulations (left panel of Fig. \ref{fig:phase-diagram}) and the photoionization emission model to characterize the impact of including the photoionization model on our results. 

The ionization parameter is defined as 
\begin{equation}
    \xi\equiv\frac{4\pi F_\mathrm{1-1000Ry}}{n_\mathrm{H}},
\end{equation}
where $F_\mathrm{1-1000Ry}$ is the local flux in the 1--1000 Ry energy band (1 Ry = 13.6 eV) following the convention of \citet{Krolik1981}. We adopted the \citet{Faucher-Giguere2020} \ac{uvb} spectral energy distribution for the calculation. We obtained a relation of $\log\xi[\mathrm{erg\ s^{-1}\ cm^2}]=2.26-\log\Delta_\mathrm{b}$ at $z=0.48$. We used the \texttt{pion} model \citep{Miller2015,Mehdipour2016} in SPEX-3.08.00 \citep{Kaastra1996,Spex3p08} for calculating the photoioized emission spectra and energy loss rates due to radiation at different $\xi$ and temperature. We note that by setting the parameter \texttt{tmod=1}, the \texttt{pion} model in SPEX allows for calculations to be carried out at fixed temperatures and ionization parameters without solving the energy balance, which meets the physical condition of \ac{whim}, whose temperature is higher than the photoionization equilibrium temperature and is maintained by external heating mechanisms such as shock heating.

\begin{figure}
    \centering
    \includegraphics[width=0.4\textwidth]{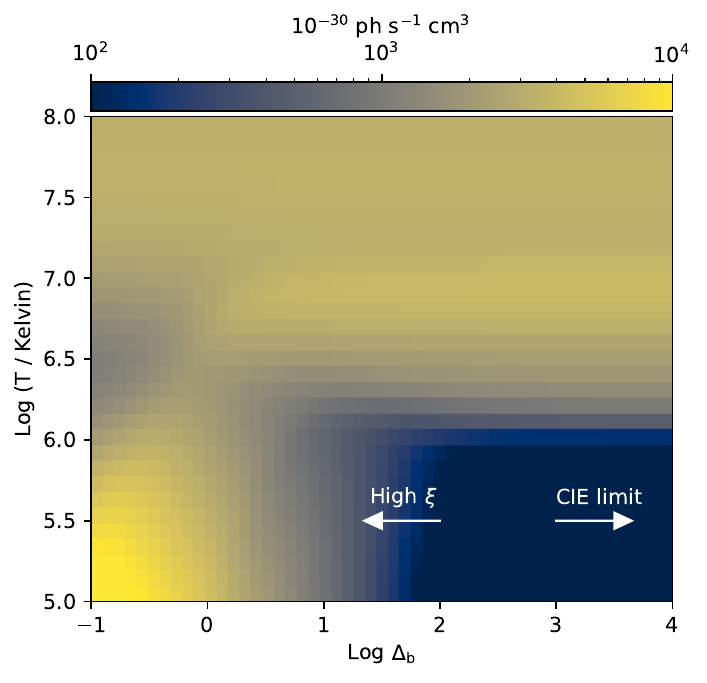}\\
    \includegraphics[width=0.4\textwidth]{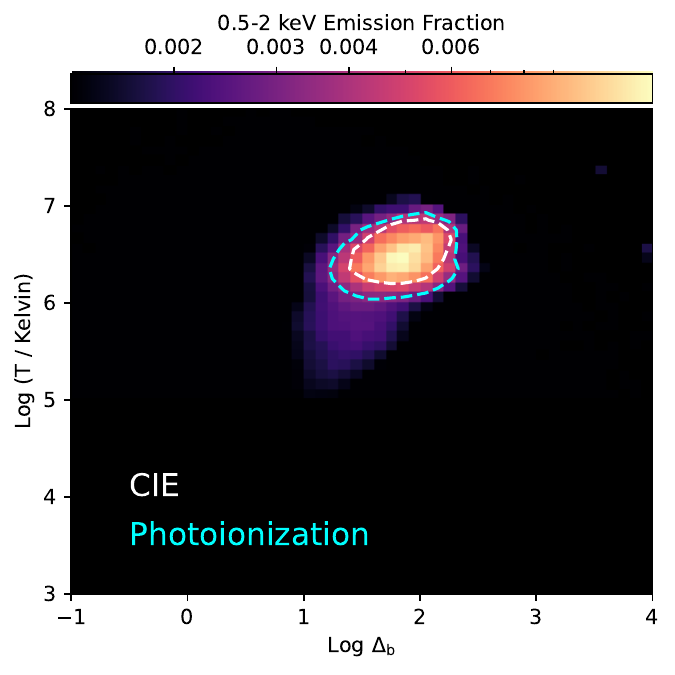}
    \caption{Impact of the cosmic UV background photoionization on the X-ray detectable phases. 
    \emph{Top}: Photon luminosity per unit of emission measure in the 0.5--2~keV band of the photoionized \ac{whim} at different density (x-axis) and temperature (y-axis) by assuming $Z=0.2Z_\sun$. Towards the high density (low ionization parameter) end, the radiative cooling reaches the \ac{cie} limit and is simply a function of temperature. When $\Delta_\mathrm{b}\lesssim1$, the radiative cooling keeps high at low temperature. 
    \emph{Bottom}: Same as the right panel of Fig. \ref{fig:phase-diagram} but calculated using the photoionization model. By taking the cosmic UV background photoionization into account, the size of the 50\% enclosed phases contour is marginally larger than that of the \ac{cie} condition. }
    \label{fig:cf_pion}
\end{figure}

The top panel of the Fig. \ref{fig:cf_pion} shows the 0.5--2~keV photon luminosity per unit of emission measure of the \ac{uvb} photonionized \ac{whim} at different temperature and density contrast. When $\Delta_\mathrm{b}>2$ ($\log\xi[\mathrm{erg\ s^{-1}\ cm^2}]\lesssim0)$, the radiative cooling reaches the \ac{cie} limit and simply becomes a function of temperature. When $\Delta_\mathrm{b}\lesssim1$ ($\log\xi[\mathrm{erg\ s^{-1}\ cm^2}]\gtrsim1)$, the radiative cooling deviates from the \ac{cie} condition and is a function of both temperature and density. The radiative cooling in the 0.5--2~keV band remains high for the \ac{whim}, with a temperature below $10^6$~K. We use the results of the photon luminosity of the photoionized \ac{whim} to update the phase diagram weighted by the 0.5--2~keV emission and plot it in the bottom panel of Fig. \ref{fig:cf_pion}. The 50\% enclosed phases contour is at the same position as that of the \ac{cie} condition with a slightly larger size. It means that our conclusion drawn from Sect. \ref{sect:phase} does not change by taking photonionization into account. The traceable \ac{whim} by broadband X-ray emission is seen to be in the highest temperature and density phases among the whole filament gas population.

\section{Summary}\label{sect:summary}
Overall, \ac{erass} has provided the deepest All-Sky X-ray survey in the soft band, allowing for a more precise investigation of X-ray emission in cosmic filaments than in previous pioneering works \citep{Tanimura2020-xray, Tanimura2022}. This work aims to use the eRASS:4 X-ray data and \ac{sdss} optical filament catalog to stack X-ray emission from filaments and study the properties of the X-ray emitting \ac{whim}. We adopted a novel method to: 1) estimate the fraction of non-WHIM emission in the stacked signal; 2) improve the spectral stacking method to minimize the systematics from the instrumental and sky background features at the observer frame; and 3) combine the \ac{sz} stacking results to simultaneously quantify the volume fraction and density of the X-ray emitting \ac{whim}. 
We summarize our main results as follows:
\begin{enumerate}
    \item We stacked 0.3--1.2~keV surface brightness profiles of 7817 \ac{sdss} optical filaments with physical lengths of $20-100$~Mpc and redshifts of 0.2--0.6 in a 2275~$\deg^2$ footprint. We detected an X-ray emission excess over the background at a $9\sigma$ significance.
    \item We validated our profile stacking method by applying it to simulated count maps, including emission from filaments with an injected model and randomized X-ray background based on the observed angular power spectrum. After 200 realizations, the recovered averaged filament profiles agree well with the injected model with only a 13\% difference, which could be the result of a possible boosting from projectively overlapped filaments in the same orientations. 
    \item We carefully estimate the fraction of stacked signal from unmasked sources, including X-ray halos, \acp{agn}, and \acp{xrb} that are associated with galaxies. We first paint the observed galaxies in the analysis footprint with X-ray count rate from $L_\mathrm{X}-M_*$ scaling relations to create an expected X-ray emission map from galaxies, where the incompleteness and bias of observed galaxy \ac{smf} is corrected. Then we apply the same surface brightness profile stacking method to obtain the fractions of stacked signal from each type of unmasked sources, which is 20\%, 12\%, and 5\% for X-ray halos, \acp{agn}, and \acp{xrb}, respectively. The photometric redshift uncertainties of the galaxies we used could underestimate another 3\% of the contamination fraction. Based on these fractions, we conclude that the remaining $\sim60\%$ of the signal is from \ac{whim} in the filaments and the \ac{whim} detection significance is $5.4\sigma$.
    \item We stacked broadband spectra and response files of the filaments in the rest frame. We included the components of unmasked sources in spectral fitting. The best-fit temperature using a \ac{1t} model is $0.58\pm0.10$~keV, which could be biased to the high end of the \ac{whim} temperature distribution given its multi-temperature nature suggested by numerical simulations. The stacked spectra can also be fitted by a numerical simulation-motivated model with a log-normal temperature distribution, which will reduce the residual in the 0.5--0.6~keV energy range, suggesting the presence of the OVII abundant temperature phases.
    \item We modeled the baryon density contrast $\Delta_\mathrm{b}$ of the X-ray emitting \ac{whim} by fitting the stacked surface brightness profile, where we additionally introduced a parameter of volume filling factor $f_\mathrm{vol}$, which is necessary because only a small fraction of \ac{whim} emits X-rays that can be observed by eROSITA. We used the \ac{sz} stacking results in the literature to break the degeneracy between the $\Delta_\mathrm{b}$ and volume fraction $f_\mathrm{vol}$ and obtained an averaged $\log\Delta_\mathrm{b}=1.88\pm0.18$, suggesting that the observed emission is from the densest population of the \ac{whim}.
    \item We analyzed the $z=0.48$ snapshot of the TNG300-1 simulation to investigate the X-ray emitting phases in simulations. We found that most of the photons in the 0.5--2~keV energy range are emitted by phases of $6.2<\log(T/\mathrm{K})<6.9$ and $1.2<\log\Delta_\mathrm{b}<2.2$, which is in general agreement with our estimated values.
\end{enumerate}

This is the first time we detected $>5\sigma$ X-ray emission from cosmic filaments. We applied a thorough investigation of the properties of the X-ray emitting \ac{whim}. In the coming decade, our understanding of X-ray emissions from filaments and the WHIM within them will significantly improve from multiple aspects. The 4-metre Multi-Object Spectroscopic Telescope (4MOST) Cosmology Redshift Survey \citep{Richard2019} will bring deep and wide field galaxy spectroscopic redshift data in the southern hemisphere, allowing us to take full use of the \ac{erass} data of the German consortium. Improved filament finders such as the Monte Carlo Physarum Machine \citep{Elek2022} and 1-DREAM \citep{Canducci2022} will be applied to new survey data and provide a more accurately constructed filament catalog. Meanwhile, a better understanding of X-ray properties of galaxy groups, \acp{agn}, and \acp{xrb} will help us to obtain a better estimate of the contamination fractions of unmasked sources. Moreover, high spectral resolution X-ray missions such as {Hot Universe Baryon Surveyor} \citep{Cui2020} and {Line Emission Mapper} \citep{Kraft2022} will be able to explore a wider parameter space of the \ac{whim} properties.

\begin{acknowledgements}
We acknowledge Jelle Kaastra for the instruction on using the SPEX Pion model. We also acknowledge the anonymous referee for the insightful comments that improved the manuscript. 
X.Z., E.B., V.G., A.L., C.G., and S.Z. acknowledge financial support from the European Research Council (ERC) Consolidator Grant under the European Union’s Horizon 2020 research and innovation program (grant agreement CoG DarkQuest No 101002585). M.C.H.Y. and M.F. acknowledge support from the Deutsche Forschungsgemeinschaft through the grant FR 1691/2-1. Y.Z. and G.P. acknowledge funding from the European Research Council (ERC) under the European Union’s Horizon 2020 research and innovation programme (grant agreement No 865637). G.P. acknowledges support from Bando per il Finanziamento della Ricerca Fondamentale 2022 dell’Istituto Nazionale di Astrofisica (INAF): GO Large program and from the Framework per l’Attrazione e il Rafforzamento delle Eccellenze (FARE) per la ricerca in Italia (R20L5S39T9). M.B. acknowledges funding by the Deutsche Forschungsgemeinschaft under Germany's Excellence Strategy -- EXC 2121 ``Quantum Universe'' --  390833306. A.V. acknowledges funding by the Deutsche Forschungsgemeinschaft -- 450861021. N. M. acknowledges funding by the European Union through a Marie Sk{\l}odowska-Curie Action Postdoctoral Fellowship (Grant Agreement: 101061448, project: MEMORY). Views and opinions expressed are however those of the author only and do not necessarily reflect those of the European Union or of the Research Executive Agency. Neither the European Union nor the granting authority can be held responsible for them. This work is based on data from \ac{erosita}, the soft X-ray instrument aboard SRG, a joint Russian-German science mission supported by the Russian Space Agency (Roskosmos), in the interests of the Russian Academy of Sciences represented by its Space Research Institute (IKI), and the Deutsches Zentrum f{\"{u}}r Luft und Raumfahrt (DLR). The SRG spacecraft was built by Lavochkin Association (NPOL) and its subcontractors and is operated by NPOL with support from the Max Planck Institute for Extraterrestrial Physics (MPE).
The development and construction of the eROSITA X-ray instrument were led by MPE, with contributions from the Dr. Karl Remeis Observatory Bamberg \& ECAP (FAU Erlangen-Nuernberg), the University of Hamburg Observatory, the Leibniz Institute for Astrophysics Potsdam (AIP), and the Institute for Astronomy and Astrophysics of the University of T{\"{u}}bingen, with the support of DLR and the Max Planck Society. The Argelander Institute for Astronomy of the University of Bonn and the Ludwig Maximilians Universit{\"{a}}t Munich also participated in the science preparation for eROSITA. 
The eROSITA data shown here were processed using the eSASS/NRTA software system developed by the German eROSITA consortium.
\end{acknowledgements}

\bibliographystyle{aa}
\bibliography{aanda}

\begin{appendix}

\section{Surface brightness profiles in different bands and length populations}\label{app:profile_more}

\begin{figure*}
    \sidecaption
    \includegraphics[width=12cm]{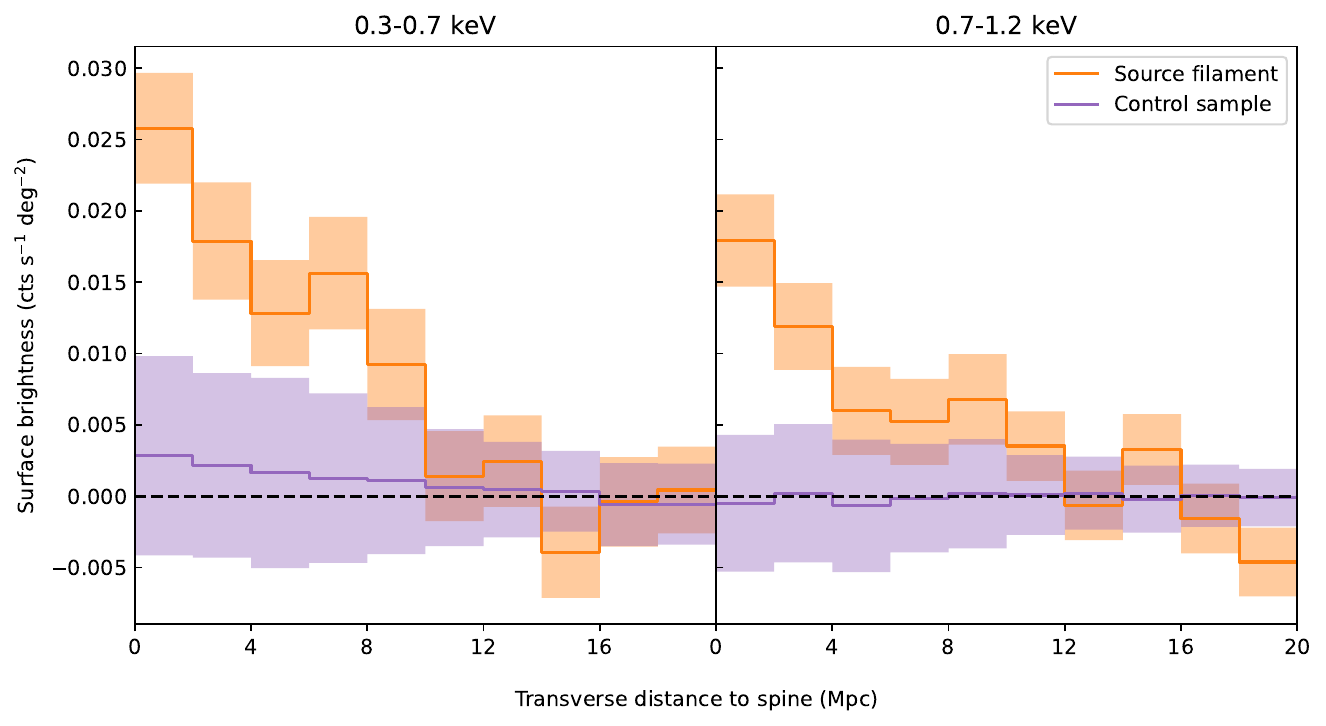}
    \caption{Same as Fig. \ref{fig:total_prof}, but in the 0.3--0.7~keV (\emph{left}) and 0.7--1.2~keV (\emph{right}) bands, respectively. }\label{fig:prof_soft-hard}
\end{figure*}

In Fig. \ref{fig:prof_soft-hard}, we plot the stacked net profile in the 0.3--0.7~keV and 0.7--1.2~keV bands. The detections in the two bands are both significant, reaching $7.0\sigma$ and $6.5\sigma$ levels, respectively. 
In contrast to previous works of \citet{Tanimura2020-xray,Tanimura2022}, we observe significant X-ray emission in the <0.6~keV energy band.

\begin{figure}
    \centering
    \includegraphics[width=0.5\textwidth]{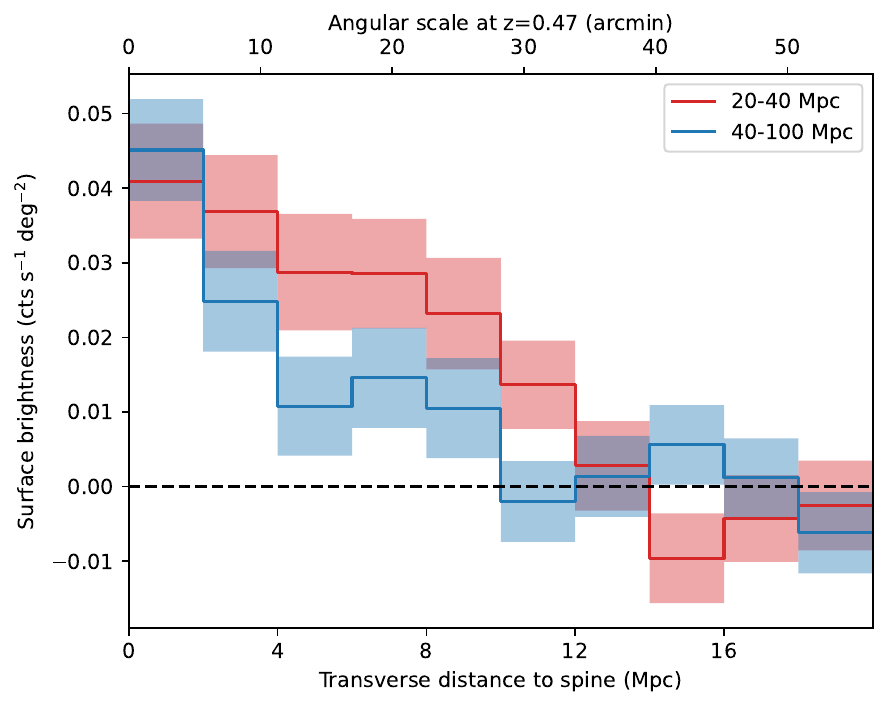}
    \caption{Local background-subtracted surface brightness profiles of the 20--40~Mpc length (red) and the 40--100~Mpc length (blue) filament samples. The dashed horizontal line denotes the zero level. }
    \label{fig:prof_short-long}
\end{figure}

We also split the filament sample into 20--40~Mpc and 40--100~Mpc length populations and we detect a marginal $2.5\sigma$ discrepancy between the two populations (see Fig.~\ref{fig:prof_short-long}). The short filaments have a broader stacked emission profile than long filaments, consistent with the predictions in numerical simulations \citep{Galarraga-Espinosa2021}. 

\clearpage

\section{Impact of the source masking radius}\label{app:mask_radius}

The choice of source masking radius could lead to a change in the stacked profile. \citet{Tanimura2022} has investigated the impact of different cluster mask radii and point source mask radii on the results. They discovered that the increase in the cluster mask radii from $r_{500}$ to $3r_{500}$ and the increase in the point source mask radii from $30\arcsec$ to $45\arcsec$ do not affect the results (within the uncertainties). In this section, we present our independent check of this systematic uncertainty in our data. 

\begin{figure*}
    \centering
    \begin{tabular}{cc}
    \includegraphics[width=0.45\textwidth]{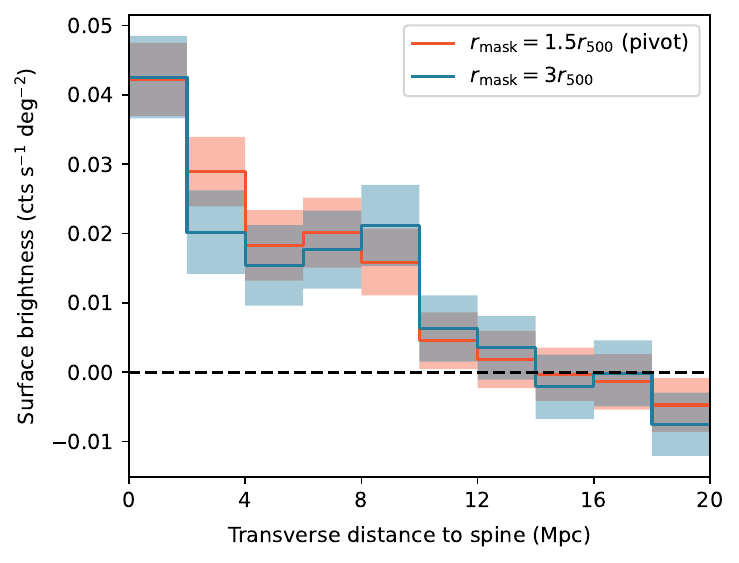} & \includegraphics[width=0.45\textwidth]{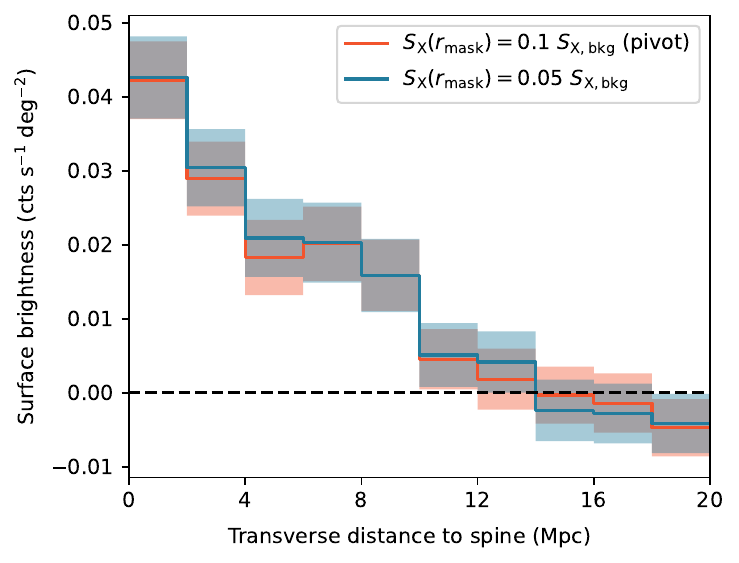}
    \end{tabular}
    \caption{Impact of the source masking radius on the stacked profile. 
    \emph{Left}: Comparison between stacked net profile using $1.5r_{500}$ (red) and $3r_{500}$ (blue) cluster mask radii. The stacked profile with large cluster masks agrees with the original profile within the uncertainties. 
    \emph{Right}: Comparison between stacked net profiles using X-ray source masks with $r|_{S=0.1S_\mathrm{bkg}}$ (red) and $r|_{S=0.05S_\mathrm{bkg}}$ (blue). A more conservative choice of the source mask radius does not change the result within uncertainties.}
    \label{fig:prof_mask_size}
\end{figure*}

\begin{figure}
    \centering
    \includegraphics[width=0.5\textwidth]{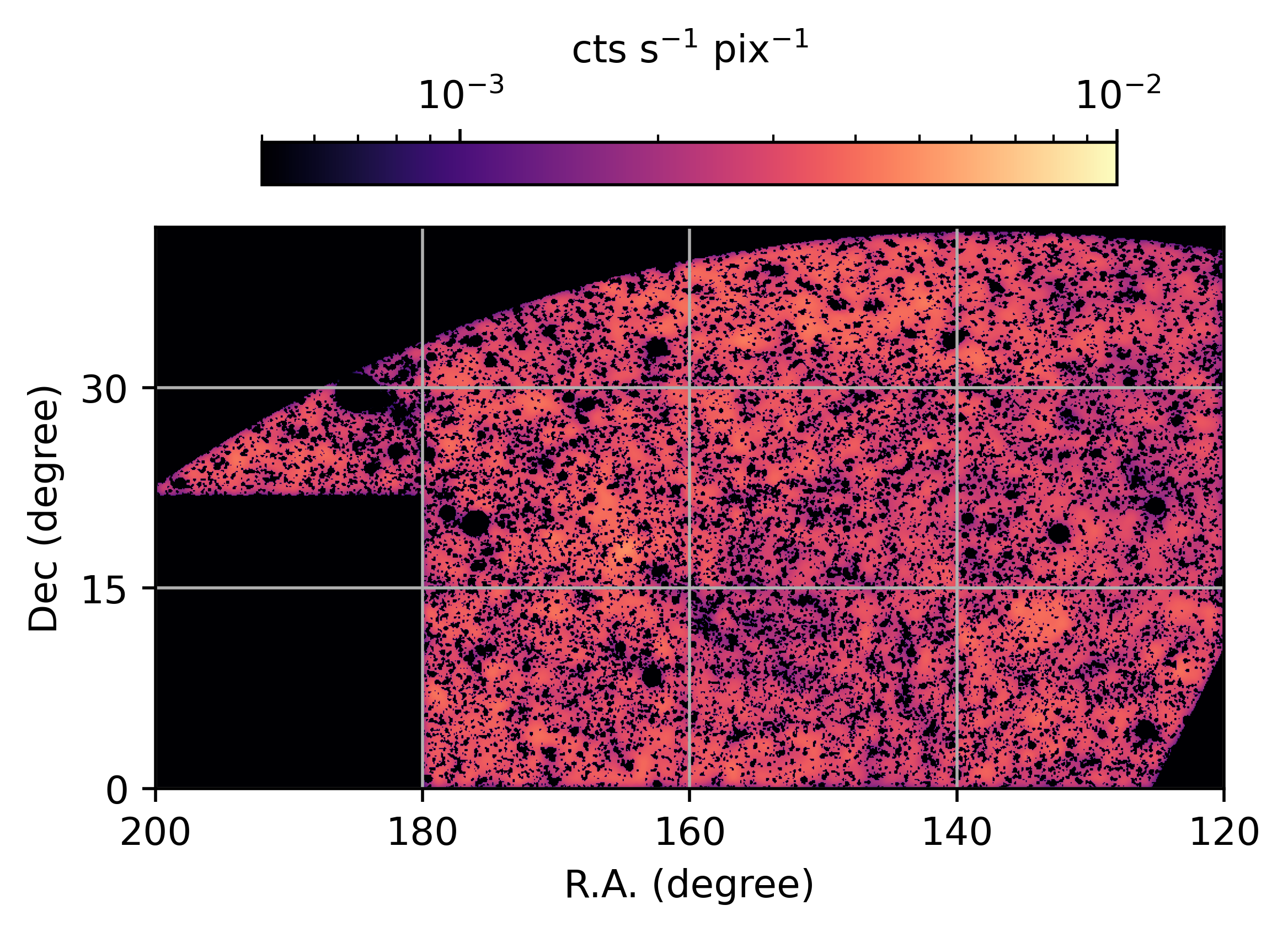}
    \caption{Same as Fig. \ref{fig:footprint}, but with $3r_{500}$ ($3r_\lambda$) mask radii for X-ray selected (optically identified) clusters. The analysis footprint without source masks is significantly reduced.}
    \label{fig:footprint_3r500}
\end{figure}

For cluster masks, instead of the $1.5r_{500}$ for the eRASS1 X-ray selected sample and $1.5r_\lambda$ for the redMaPPer sample (as described in Sect. \ref{sect:mask}), we increased the masking radii by a factor of two for a test. Following Sects. \ref{sect:mask} and \ref{sect:healpix}, we create new 0.3--1.2~keV count and exposure Healpix maps with clusters masked using $3r_{500}$ and $3r_\lambda$ masks. Other source masks are the same as those listed in Table~\ref{tab:source_mask}. In the left panel of Fig.~\ref{fig:prof_mask_size}, we plot the stacked net profiles with the doubled cluster masking radii. The profile with doubled mask radii agrees with the original one within uncertainties. The integrated net signal within 10~Mpc transverse distance is 93\% of the original integrated signal. We note that the uncertainty in each bin is increased by $13\%$, which is due to the significant reduction of the available footprint (see the comparison between Fig.~\ref{fig:footprint} and Fig.~\ref{fig:footprint_3r500}).

Similarly, for the eRASS:4 X-ray sources, we tested on the masking radii that are calculated as $S(r_\mathrm{cut})=0.05S_\mathrm{bkg}$ and plot the comparison in the right panel of Fig. \ref{fig:prof_mask_size}. The profiles using two different mask radii are identical within the uncertainties.

\clearpage

\section{Mock maps for profile stacking validation}\label{app:mock_map}

\begin{figure*}
    \centering
    \begin{tabular}{cc}
        \includegraphics[width=0.4\textwidth]{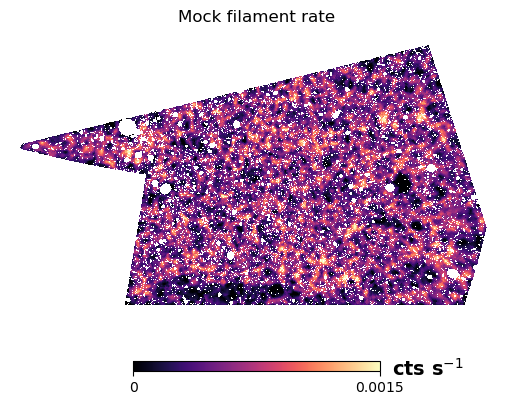}&
        \includegraphics[width=0.4\textwidth]{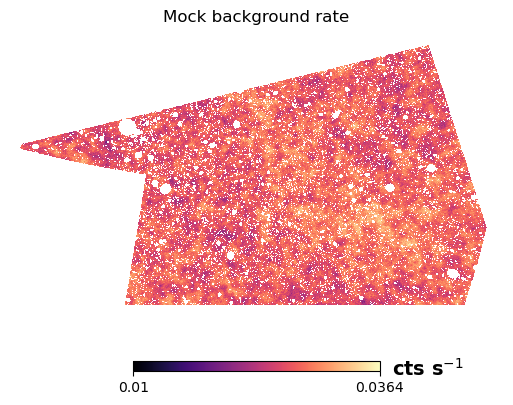}\\
        \includegraphics[width=0.4\textwidth]{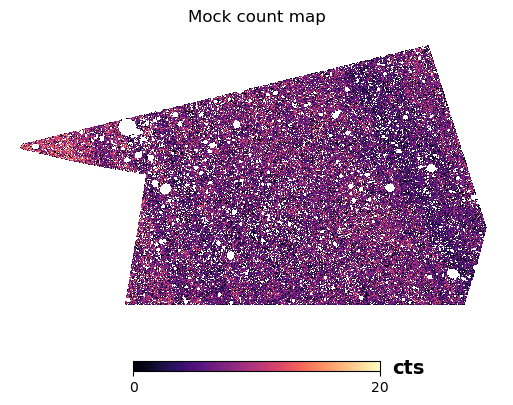}&
        \includegraphics[width=0.4\textwidth]{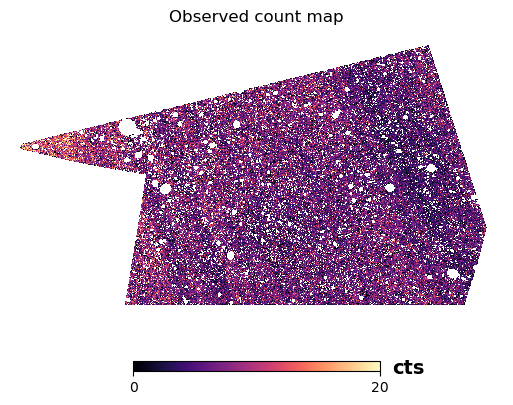}
    \end{tabular}
    \caption{Example of simulated maps in the 0.3--1.2~keV band for validating the imaging stacking method. \emph{Top left}: Mock count rate map for filament emission. \emph{Top right}: Mock foreground/background emission mock map. \emph{Bottom left}: Mock count map of the total emission. \emph{Bottom right}: Observed count map as a comparison.
    }
    \label{fig:mock_map}
\end{figure*}

First, we phenomenically fit the stacked net surface brightness profile using a King-profile model (see the section on modeling gas properties in the filaments). With the best-fit parameters, we created the mock filament emission rate map based on the projected geometric configurations of the 7817 filaments (top left panel of Fig.~\ref{fig:mock_map}). We note that the best-fit parameters in this section are empirical and we did not use them for physical interpretation. The goal is to test the methodology and algorithms used to extract stacked surface brightness profiles. The best-fit profile for injection is plotted as the dotted line in Fig.~\ref{fig:mock_profile}.

We simulated mock background rate maps based on the angular power spectrum of the observed rate map (top right panel of Fig.~\ref{fig:mock_map}). By comparing the mock filament emission map and the observed sky map, we know that the filament emission contributes $<1\%$ of the total signal. Therefore, we assume that the angular power spectrum of the observed rate map adequately represents that of the background component. We first smooth the rate map using a $1^{\circ}$ \ac{fwhm} Gaussian kernel to reduce the Poisson noise. Then, we construct a relative amplitude map based on the mean and standard deviation of the smoothed rate map. We computed the angular power spectrum of the relative amplitude map $C_{l,\mathrm{rel}}$ up to a maximum multipole moment $l=3000$ using the function \texttt{anafast} in the HEALPY package. We simulated a mock relative amplitude map using this angular power spectrum and convert it to a mock rate map using the mean and standard deviation of the smoothed rate map. 
We summed up the mock filament rate map and the mock background rate map to get a mock sky rate map. We obtained a mock sky count map by multiplying the exposure map from observations and applying Poisson randomization to the expected count value in each pixel. 

\clearpage

\section{Spectral models for undetected galactic sources}

\begin{figure*}
    \centering
    \includegraphics[width=.99\textwidth]{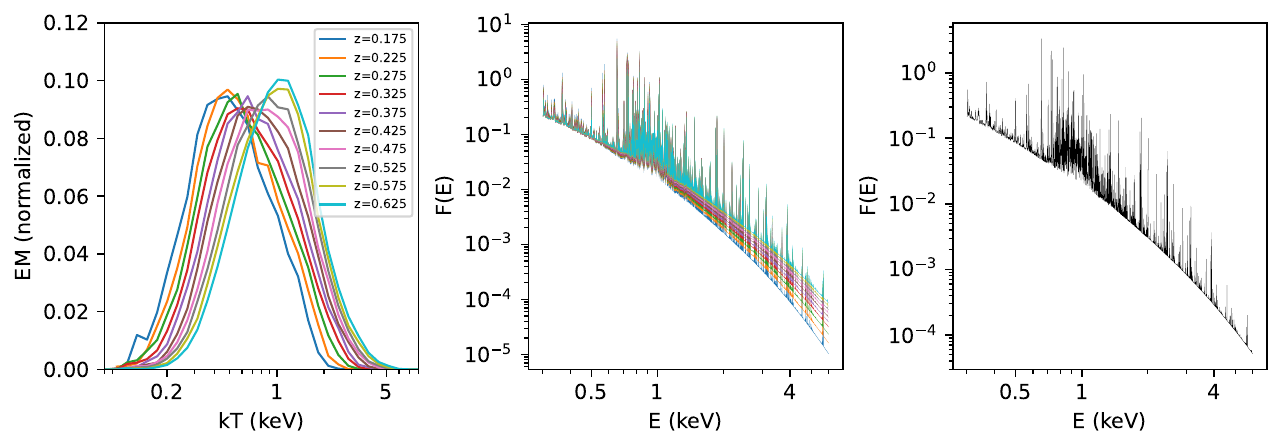}
    \caption{\emph{Left}: Multi-temperature differential emission measure distribution of undetected X-ray halos at different redshifts after applying the selection. \emph{Middle}: Corresponding composite spectra of each differential emission measure distribution at different redshifts. \emph{Right}: Synthesis rest-frame spectral model of undetected X-ray halos by taking the filament sample length and redshift distributions into account. }
    \label{fig:spec-dem}
\end{figure*}

\begin{figure}
    \centering
    \includegraphics[width=0.5\textwidth]{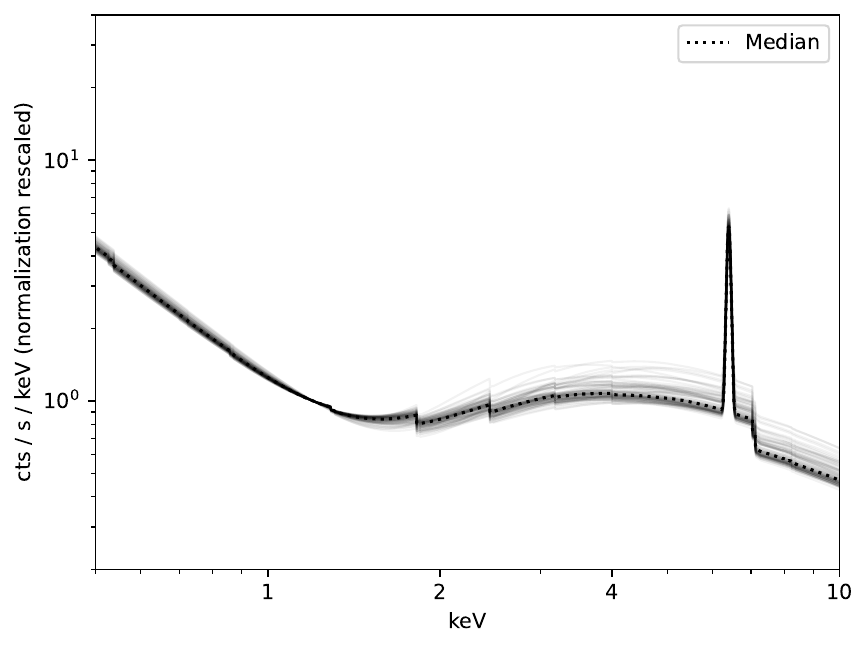}
    \caption{Individual undetected AGN spectral models in different redshift - stellar mass bins of the mock AGN catalog with the normalizations rescaled. The spectra show strong self-similarity in the soft band. The dotted is the median of all the spectra. }
    \label{fig:agn-spec}
\end{figure}

In Fig. \ref{fig:spec-dem}, we plot the emission measure distribution of unmasked halos at different redshifts (left panel), the composite spectra of unmasked halos at different redshifts (middle panel), and the averaged spectrum (right panel). In Fig. \ref{fig:agn-spec}, we plot the spectral models in different redshift bins, stellar mass bins used in the mock catalog, and the median spectrum. 

\end{appendix}

\end{document}